\documentclass[conference]{IEEEtran}

\IEEEoverridecommandlockouts
\usepackage{cite}
\usepackage{amsmath,amssymb,amsfonts}
\usepackage{algorithmic}
\usepackage{graphicx}
\usepackage{textcomp}
\usepackage{xcolor}
\usepackage{fancyhdr}
\usepackage{fancyhdr}
\usepackage{hyperref}
\usepackage{algorithm}
\usepackage{algorithmic}
\usepackage{xspace}
\usepackage[caption=false]{subfig}
\usepackage{amsmath}
\usepackage{mathtools}
\usepackage{xcolor}
\usepackage{amsfonts}
\usepackage{amssymb}
\usepackage{pifont}
\usepackage{color}
\usepackage{multirow}
\usepackage{soul} 
\usepackage{tikz}

\usepackage{color, xcolor,colortbl}
\definecolor{green}{rgb}{0.1,0.1,0.1}

\usepackage{bbding}
\usepackage{pifont}
\usepackage{wasysym}
\usepackage{amssymb}

\newcommand{\ignore}[1]{}

\newcommand{\proposed}[0]{\texttt{Centaur}\xspace}
\newcommand{\streamer}[0]{\texttt{EB-Streamer}\xspace}
\newcommand{\bpr}[0]{\texttt{BPregs}\xspace}
\newcommand{\ebgu}[0]{\texttt{EB-GU}\xspace}
\newcommand{\ebru}[0]{\texttt{EB-RU}\xspace}
\newcommand{\spid}[0]{\texttt{SRAM$_{sparseID}$}\xspace}
\newcommand{\mlpmodel}[0]{\texttt{SRAM$_{MLPmodel}$}\xspace}
\newcommand{\mlpinput}[0]{\texttt{SRAM$_{MLPinput}$}\xspace}
\newcommand{\densefeature}[0]{\texttt{SRAM$_{DenseFeature}$}\xspace}

\newcommand{\old}[1]{}
\newcommand{\fig}[1]{Figure~\ref{#1}}
\newcommand{\sect}[1]{Section~\ref{#1}}
\newcommand{\tab}[1]{Table~\ref{#1}}

\newcommand{\drain}[1]{\texttt{DRAIN}\xspace}

\newcommand{\cpuonly}[0]{\texttt{CPU-only}\xspace}

\newcommand{\cpugpu}[0]{\texttt{CPU-GPU}\xspace}

\newcommand{\dlrm}[1]{DLRM({#1})\xspace}

\usepackage{cite}
\usepackage{amsmath,amssymb,amsfonts}
\usepackage{algorithmic}
\usepackage{graphicx}
\usepackage{textcomp}
\usepackage{xcolor}
\def\BibTeX{{\rm B\kern-.05em{\sc i\kern-.025em b}\kern-.08em
    T\kern-.1667em\lower.7ex\hbox{E}\kern-.125emX}}

\begin{document}
\title{Centaur: A Chiplet-based, Hybrid Sparse-Dense\\Accelerator for Personalized Recommendations\\
}

\author{\IEEEauthorblockN{Ranggi Hwang\qquad Taehun Kim\qquad Youngeun Kwon\qquad
Minsoo Rhu}
\IEEEauthorblockA{School of Electrical Engineering\\
KAIST\\
\{\texttt{ranggi.hwang, taehun.kim, yekwon, mrhu\}@kaist.ac.kr}}}
\maketitle
\newcommand\blfootnote[1]{%
\begingroup
\renewcommand\thefootnote{}\footnote{#1}%
\addtocounter{footnote}{-1}%
\endgroup
}

\maketitle

\pagestyle{empty}

\linespread{0.98}

\begin{abstract}

	Personalized recommendations are the backbone machine learning (ML)
		algorithm that powers several important application domains (e.g., ads,
				e-commerce, etc) serviced from cloud datacenters.  Sparse embedding
		layers are a crucial building block in designing recommendations
		yet little attention has been paid in properly accelerating this
		important ML algorithm. This paper first provides a detailed workload
		characterization on personalized recommendations and
		identifies two significant performance limiters: memory-intensive embedding
		layers and compute-intensive multi-layer perceptron (MLP) layers. We then
		present Centaur, a chiplet-based hybrid sparse-dense accelerator that addresses both
		the memory throughput challenges of embedding layers and the compute
		limitations of MLP layers. We implement and demonstrate our proposal on an
		Intel HARPv2, a package-integrated CPU+FPGA device, which shows a
	    1.7$-$17.2$\times$ performance speedup and 1.7$-$19.5$\times$
		energy-efficiency improvement than conventional approaches.

	 \end{abstract}


\blfootnote{
This is the author preprint version of the work. The authoritative version will appear in the 
	Proceedings of the $47^{\text{th}}$ IEEE/ACM International Symposium on Computer Architecture (ISCA-47), 2020.
}

\section {Introduction}
\label{sect:intro}

The complexity of deep neural network (DNN) based machine learning (ML)
	algorithms is scaling up rapidly. As such, GPUs or ASIC/FPGA-based 
	ML accelerators  are
	widely being adopted for accelerating the	computationally \emph{dense} DNN
	layers. Examples include convolutional neural networks (CNNs), recurrent
	neural networks (RNNs), and multi-layer perceptrons (MLPs), all of which are
	amenable for hardware acceleration thanks to their highly regular and
	deterministic dataflow.

\begin{figure}[t!] \centering
\includegraphics[width=0.45\textwidth]{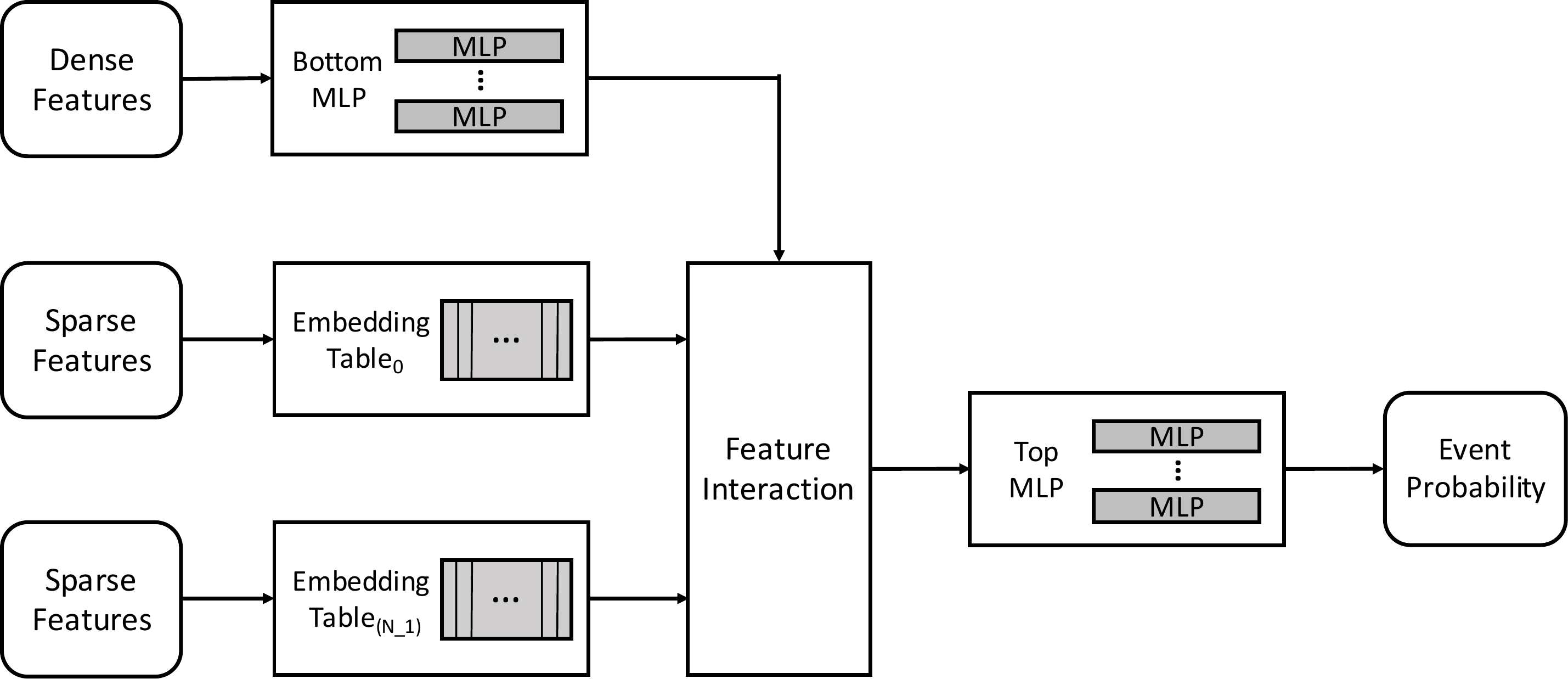}
\caption{
	Topological structure of a DNN-based personalized recommendation model
		containing sparse embedding layers as the frontend and dense DNN layers
		as the backend processing step.
}
\vspace{-0.75em}
\label{fig:upcoming_dnns}
\end{figure}

	While we were able to make significant strides in accelerating these
	compute-intensive	DNN layers, little attention has been paid in addressing
	the challenges of memory limited \emph{non}-DNN layers in emerging ML
	workloads.  Consequently, we are witnessing these non-DNN layers, especially
	those that are memory intensive, gradually becoming a more
	significant performance
	bottleneck~\cite{tensordimm,dlrm:arch,snu:ahn:batchnorm}.  In particular, ML
	algorithms employing \emph{sparse embedding layers} exhibit drastically
	different characteristics than conventional \emph{dense} DNN layers.
	\fig{fig:upcoming_dnns} illustrates the high-level structure of ML
	applications employing embedding layers, which are being adopted in a variety
	of application domains such as ads, social networking service, e-commerce,
	and others. The backbone ML algorithms that power these applications are
	\emph{personalized recommendation systems}, the most widely deployed ML
	workload serviced from the cloud.  As we study in this paper, embedding
	layers account for a significant fraction of the inference time of
	recommendations.  Consequently, several hyperscalers such as
	Google~\cite{dean:2018:goldenage},
	Facebook~\cite{park:2018:fb,dlrm:arch,sigarch:blog:recsys},
	Alibaba~\cite{kdd:alibaba}, and Baidu~\cite{hestness:2019:ppopp} all pinpoint
	to these embedding layers as causing a severe performance bottleneck in
	production-level personalized recommendation models.

In this paper, we focus on addressing the system-level challenges in
\emph{deploying} DNN-based recommendation models.  Personalized recommendation
consists of two key modules: (1) the frontend, \emph{sparse} embedding layers
and (2) the backend, \emph{dense} MLP layers.  As detailed in
\sect{sect:characterization}, embedding layers consume up to several hundreds
of GBs of memory capacity in recommendations, even for inference.
Because such requirement is far beyond the physical memory capacity of
GPUs (only available with several tens of GBs, $32$ GB in NVIDIA
		V100~\cite{volta_v100}), a common practice in deploying these  
models over the cloud is to utilize the capacity-optimized CPU memory to store
the embeddings and utilize the CPU exclusively for inference (\cpuonly)~\cite{hazelwood:2018:hpca,dlrm:arch,park:2018:fb}.

Given this landscape, this paper first conducts a workload
characterization study on state-of-the-art personalized recommendation models.
We identify the following {\bf key challenges} in deploying personalized
recommendations on conventional \cpuonly systems.  During the frontend
embedding layer stage, multiple embedding vectors are \emph{gathered} from
several embedding tables which are subsequently \emph{reduced} over the
low-bandwidth CPU memory (\fig{fig:upcoming_dnns}).  Because the aggregate size
of the gathered embeddings for inference is much smaller than the size of the embedding
tables (e.g., several KBs or MBs of read over several tens of GBs of embedding tables),
			 the embedding gather operations are extremely \emph{sparse} with low
			 spatial/temporal locality, exhibiting high last-level cache (LLC) miss
			 rates (\sect{sect:caching}). Unlike throughput-optimized
			 GPUs however, CPUs are primarily optimized for latency with only a
			 handful of concurrent threads and miss status holding registers (MSHRs).
			 As such, we observe that
			 CPUs fail to maximize memory-level parallelism thus significantly under-utilizing
			 memory bandwidth for such
			 sparse embedding gather operations (\sect{sect:mem_bw}).  
			 Consequently, these sparse embedding layers
			 can account for a significant fraction of inference time (up to $79\%$),
			 causing a performance bottleneck. Another significant challenge with
			 \cpuonly recommendations is that the compute-intensive MLPs are
			 executed using the low-throughput CPUs, experiencing significant latency
			 overheads.  Overall, we identify the \emph{limited memory throughput 
				 utilization} of CPU memory systems and the \emph{low computational
					 throughput} of CPUs as the two most significant obstacles in
					 addressing the system-level bottlenecks of personalized recommendation.

To this end, we present \proposed, a chiplet-based hybrid sparse-dense
FPGA accelerator that holistically addresses the challenges of personalized
recommendations.  FPGAs have recently had a surge of interest for ML
acceleration thanks to their power-efficient and highly programmable nature.
However, prior work on FPGA-accelerated ML primarily targets the
	compute-intensive \emph{dense} DNN
		layers~\cite{intel:2018:fpga,intel:2017:fpga,intel:2017:fpl,alwani:2016:fusedCNN,fpga:dense1,
			fpga:dense2, fpga:dense3, fpga:dense4,
			fpga:dense5,cloud_dnn,dnnbuilder,dnnweaver,tabla,cosmic}, so they cannot
			properly address the challenges of sparse embeddings.  Traditionally, the
			most commonly employed integration tier between the CPU and FPGA is to
			connect them over the PCIe I/O bus, each with its own \emph{local}
			physical memory.  The FPGA in effect functions as a discrete co-processor
			device (similar to discrete GPUs) and provides ML acceleration as a
			service to the CPU via a task \emph{offloading} request.
			Recent advances in chiplet
				technology~\cite{intel:emib,mcm_gpu,simba} however enabled a more
				tight CPU+FPGA integration at the \emph{package-level}, providing
				high-bandwidth and low-latency communication between the CPU and FPGA
				chiplets over a physically \emph{shared} memory. This allows
			the FPGA chiplet to directly access the embedding tables stored inside
			the CPU DIMMs, obviating the need for memory copy operations between host
			and I/O device memory as required in
			discrete GPUs or FPGAs.
			Furthermore,
				as these
					package-integration technology matures, we expect an even higher compute
						density as well as higher inter-chip communication
						bandwidth~\cite{intel:emib,agilex,mcm_gpu}.  The {\bf key
							innovation} of \proposed  is the utilization of this emerging,
							chiplet-based CPU+FPGA technology to develop a
							\emph{heterogeneous} accelerator architecture tailored to address
							the conflicting resource requirements of recommendation models.
							Concretely, \proposed synergistically combines the following two
							modules for high-performance recommendations:

\begin{enumerate}

\item {\bf ``Sparse'' accelerator for embeddings}: Under our package-integrated
CPU+FPGA, the FPGA can directly read (write) from (to) the shared physical
memory over the high-bandwidth, low-latency CPU$\leftrightarrow$FPGA communication links.
\proposed implements a \emph{sparse} accelerator for high-throughput embedding
\emph{gather} and \emph{reduction} operations,  directly streaming out the embeddings and
reducing them from the CPU memory. This helps improve \proposed's effective throughput in embedding
gathers and reductions, achieving
superior memory bandwidth utilization for embedding layers.

\item {\bf ``Dense''  accelerator for GEMMs}: Alongside our sparse accelerator,
	\proposed incorporates a module for accelerating the compute-intensive DNN
	layers. We design a \emph{dense} accelerator to handle the GEMM (general
			purpose matrix multiplication) operations such as MLPs or feature
	interactions, allowing significant latency reduction compared to the baseline
	\cpuonly which relies on low-throughput CPU cores for GEMM operation.

\end{enumerate}

Overall, our \proposed design utilizes the unique properties of 
package-integrated CPU+FPGAs to demonstrate the merits of a chiplet-based, hybrid
sparse-dense
accelerator architecture that effectively tackles the performance bottlenecks of
personalized recommendation. Specifically, our sparse-optimized accelerator helps overcome the
limited memory bandwidth utility of \cpuonly and achieves significant throughput improvements
for sparse embedding layers. Furthermore, \proposed
improves the performance of MLP layers thanks to the high-throughput FPGA
logic. Putting everything together, \proposed provides $1.7$$-$$17.2$$\times$ speedup
and $1.7$$-$$19.5\times$ energy-efficiency improvement
than \cpuonly in deploying end-to-end personalized
recommendation models.

\section{Background}
\label{sect:background}

\begin{figure}[t!] \centering
\includegraphics[width=0.53\textwidth]{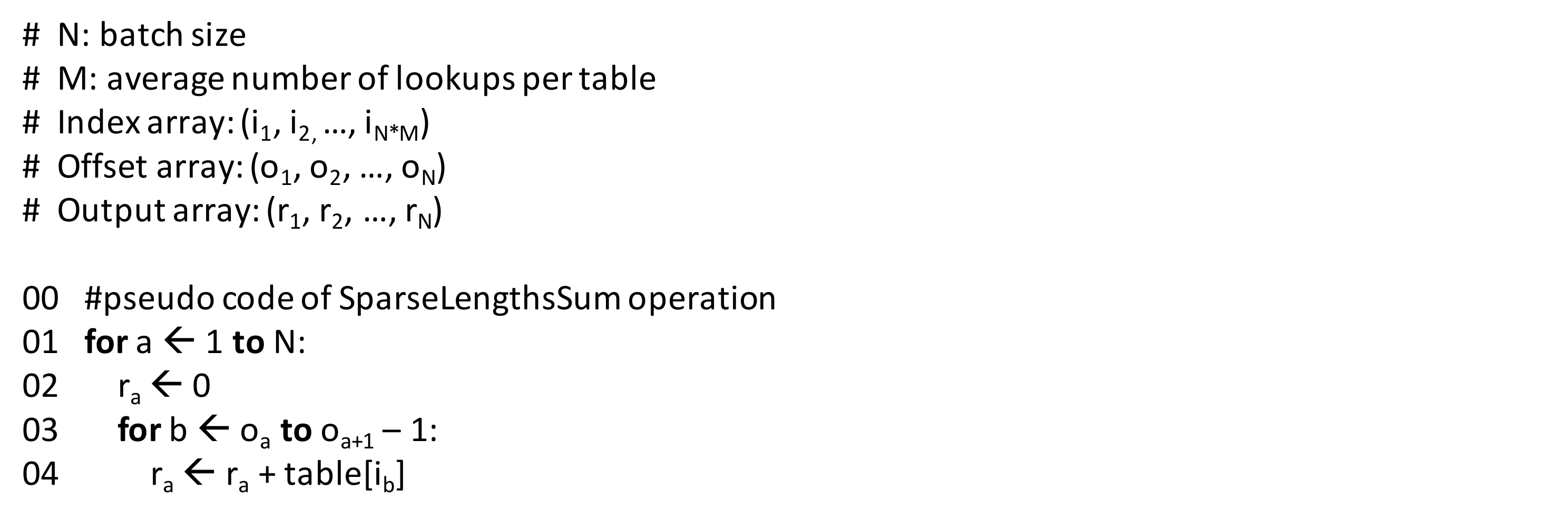}
\vspace{-1.5em}
\caption{
Functional behavior of \texttt{SparseLengthsSum()} in Caffe2~\cite{caffe2},
which conducts embedding gathers and reductions.
}
\vspace{-0.75em}
\label{fig:sls_code}
\end{figure}

\subsection{Sparse vs. Dense Layers in Personalized Recommendations}
\label{sect:layer_type}


The computer systems community has primarily focused on accelerating the
computationally intensive CNNs, RNNs, and MLPs, which exhibit a \emph{dense}
and highly \emph{regular} computational property.  Because of its highly
deterministic dataflow, these dense DNN layers are amenable for hardware
acceleration using custom-designed architectures for training and
inference~\cite{eyeriss,whatmough:2017:hotchips,park:2018:olaware,diannao,dadiannao,cambricon,song:2015:eie,scnn,cnvlutin,fpga:dense1,
	fpga:dense2, fpga:dense3,
	fpga:dense4,wu:2019:fb_edge,rhu:2016:vdnn,mcdla,mcdla:cal,kwon:2019:disagg,rhu:2018:cdma,neummu,choi:2020:prema,jang:2019:mnnfast}.

\begin{figure}[t!] \centering
	\includegraphics[width=0.485\textwidth]{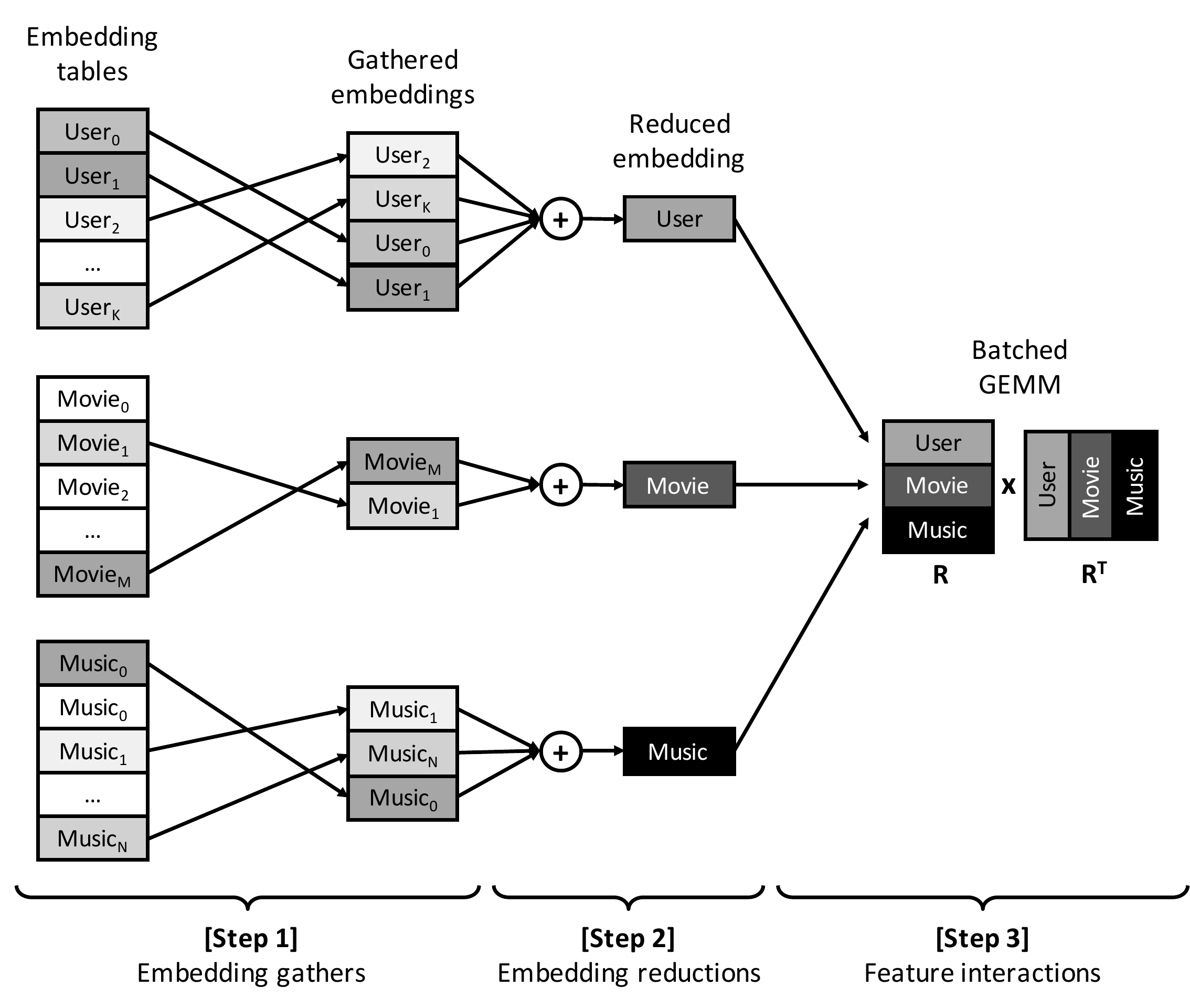}
\vspace{-1.5em}
\caption{
Illustration of embedding gather and reduction operations, followed by a feature interaction stage. The example assumes three embedding tables are used, each with $4$, $2$, and $3$ gather operations per each table. The feature interaction stage is conducted by a batched
	GEMM operation, the input of which is collected by concatenating
	the three reduced embeddings as a tensor.
}
\vspace{-0.75em}
\label{fig:sls_example}
\end{figure}

However, emerging ML	workloads employing embedding layers exhibit a highly
\emph{irregular} and \emph{sparse} dataflow.
\fig{fig:sls_code} is a pseudo-code of the
\texttt{SparseLengthsSum} function implemented in Caffe2~\cite{caffe2}, which
conducts embedding \emph{lookups} (aka \emph{gathers}) and embedding (vector)
	\emph{reductions}, widely employed in DNN-based recommendation
	systems~\cite{facebook_dlrm}. Millions of vectors called embeddings are
	stored contiguously inside a table, called embedding (lookup) table, and a
	sparse index ID is used to lookup a unique row from this table. An embedding
	gather operation takes \emph{multiple} sparse indices as inputs, which do not
	necessarily point to contiguous rows within the embedding table, to lookup
	multiple rows from this table.  Consequently, an embedding gather operation
	exhibits a highly sparse and random memory access pattern with low
	temporal/spatial locality. The embedding vectors gathered from the lookup
	table can be combined with other vectors using element-wise
	(addition/multiplication/$\ldots$) operations, hence performing reductions as
	illustrated in \fig{fig:sls_example}. The reduced embedding vectors go through
	a feature interaction step to algorithmically capture the complex interaction
	between different embedding features. While several implementations exists for
	feature interaction~\cite{facebook_dlrm}, we assume the dot-product based
	feature interaction method as employed in Facebook's open-sourced deep learning recommendation
			model (DLRM)~\cite{facebook_dlrm}. The feature interaction stage in DLRM is 
implemented by taking the dot-product between all pairs of (reduced) embedding vectors (the batched
		GEMM operation in \fig{fig:sls_example}), the outputs of which are all concatenated 
with the output vector of the bottom MLP layer (\fig{fig:upcoming_dnns}). The concatenated vector is
then post-processed with the top MLP and fed into a Sigmoid function to calculate an event probability (e.g.,
		the likelihood of a Facebook user clicking an advertisement banner).

\subsection{ML Workloads using Embeddings}
\label{sect:emb_apps}

An embedding is a projection of a discrete, categorical feature into a vector
of continuous real numbers.  Under the context of our ML workloads, embeddings
are low-dimensional, learned vector representations of feature variables, which
have recently shown to be very effective in numerous application domains such
as recommendation systems~\cite{he:www:2017,facebook_dlrm,kdd:alibaba}, machine
translation~\cite{bert}, and automatic speech recognition~\cite{deepspeech_2}.
A recommendation system for instance is formulated as a problem of
estimating the likelihood of a certain event. A DNN-based recommendation is designed
to utilize embeddings to take into account each user and item's learned
features and use embedding reductions to interact different features
altogether, which is later processed by a backend DNN execution step to extract
the probability of a certain event. 

\begin{figure}[t!] \centering
\includegraphics[width=0.485\textwidth]{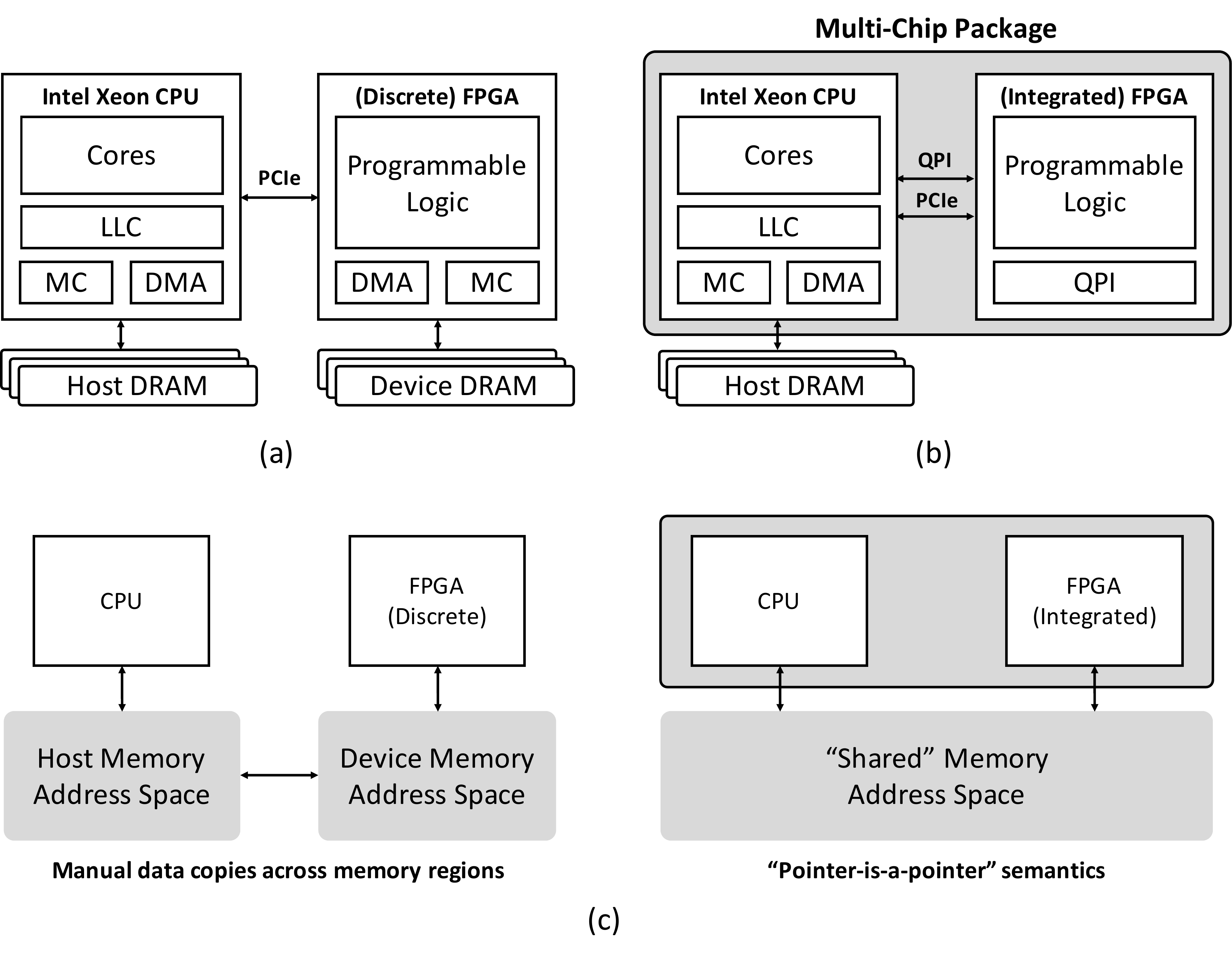}
\vspace{-1.5em}
\caption{
CPU$\leftrightarrow$FPGA integration tiers assuming (a) a \emph{discrete} FPGA communicating with the CPU over
the	PCIe I/O bus, and (b) a \emph{package-integrated} CPU+FPGA housed inside a single CPU socket. (c) The package-level
integration of CPU+FPGA enables a \emph{shared} memory address space between the CPU and FPGA, allowing high-bandwidth, low-latency
communication between the CPU and FPGA at the hardware level. As this paper utilizes
Intel's HARPv2~\cite{intel:harpv2} to demonstrate the merits of chiplet-based CPU+FPGA for recommendations,
	we assume Intel's technology (e.g., QPI) and nomenclature for the rest of this paper. Nonetheless, the high-level 
		intuitions of our proposal are equally applicable for alternative chiplet-based CPU+FPGA designs.
}
\vspace{-0.75em}
\label{fig:fpga_sys}
\end{figure}

\subsection{Discrete vs. Integrated FPGAs for ML Acceleration}
\label{sect:fpga_sys}

While ASICs provide significant energy-efficiency gains than general-purpose
CPUs/GPUs for dense DNN layers, they are not able to flexibly cope with the
ever-evolving ML algorithm research space. Reconfigurable processor
architectures such as FPGAs  represent an
intermediate design point between the efficiency of ASICs and the
programmability of general purpose (CPU/GPU) processors, providing the
potential for flexible acceleration of the constantly evolving ML
applications~\cite{intel:2018:fpga,intel:2017:fpga,intel:2017:fpl,alwani:2016:fusedCNN,fpga:dense1,
	fpga:dense2, fpga:dense3, fpga:dense4, fpga:dense5}.  The most widely
	employed CPU-FPGA integration strategy is to connect a discrete FPGA card to
	the CPU over the I/O bus (i.e., PCIe), both of which is equipped with its own local physical memory (\fig{fig:fpga_sys}(a)).  Many FPGA boards employ this
	style of integration because of its extensibility and the high throughput it
	can provide to the CPU as a co-processor device.  A key challenge with such
	integration tier is that the CPU$\leftrightarrow$FPGA communication speed is
	bounded by the narrow PCIe bus bandwidth and its high latency, so the
	benefits of FPGA acceleration is only provided when its benefits outweigh the
	task offloading overhead.  More recent products therefore employ a more tight
	CPU+FPGA integration at the \emph{package-level}, allowing the CPU and FPGA
	chiplets to communicate at a much higher bandwidth and lower latency than discrete
	FPGAs (\fig{fig:fpga_sys}(b)), with future designs expected to provide
	even higher bandwidth and speed using more advanced multi-chip
	packaging technologies~\cite{intel:emib,agilex,mcm_gpu}.
	Another key advantage of integrated CPU+FPGA
	devices is that they can share a single physical memory, which allows
	fine-grained FPGA-to-CPU data accesses  (and vice versa) at the
	hardware-level, obviating the latency overheads of traversing through the
	software stack for data movements (i.e., manual DMA-invoked \texttt{memcpy}s across the
			CPU$\leftrightarrow$FPGA memory address space, \fig{fig:fpga_sys}(c))
	thus reducing overall memory access latency.

\section{Workload Characterization of DNN-based Personalized Recommendation Systems}
\label{sect:characterization}

In this section, we utilize the open-sourced deep learning
recommendation model (DLRM)~\cite{facebook_dlrm} to conduct a
detailed workload characterization study on DNN-based personalized
recommendations. DLRM comes with several production-level model configurations
and we generate six recommendation models that covers the design space of
recommendations (as discussed in \cite{facebook_dlrm,dlrm:arch}) by
varying the number of embedding tables, number of gather operations per each
table, and the total memory usage of embedding tables and MLP layers (\tab{tab:benchmarks}).  
Following prior work~\cite{facebook_dlrm,dlrm:arch}, each embedding
is sized as a $32$-dimensional vector as default.
A key objective of our characterization study is to
root-cause the performance bottlenecks of recommendation models and motivate
our hybrid sparse-dense FPGA accelerator design. In the rest of this paper, we assume the
\cpuonly system as our baseline architecture as it is 
the most commonly deployed system design point for 
recommendations.
We further detail the merits of \cpuonly for deploying recommendations in \sect{sect:motivation}.

\begin{table}[t!]
  \centering
  \caption{Recommendation model configurations.}
\scriptsize
  \begin{tabular}{|c|c|c|c|c|}
    \hline
    \textbf{Model} & \textbf{\# of Tables} & \textbf{Gathers/table}& \textbf{Table size} & \textbf{MLP size} \\
    \hline
    \hline
    \texttt{DLRM(1)}						&		$5$		& $20$	 & $128$ MB		& $57.4$ KB\\
    \hline	                                                       
		\texttt{DLRM(2)}						&		$50$	& $20$	 & $1.28$ GB	& $57.4$ KB\\
    \hline                                                         
    \texttt{DLRM(3)}   					&   $5$		& $80$	 & $128$ MB		& $57.4$ KB\\
    \hline                                                         
    \texttt{DLRM(4)}   					&   $50$	& $80$		& $1.28$ GB	& $57.4$ KB\\
    \hline                                                         
    \texttt{DLRM(5)}   					&   $50$	& $80$		& $3.2$ GB	& $57.4$ KB\\
    \hline                                                         
    \texttt{DLRM(6)}   					&   $5$		& $2$			& $128$ MB	& $557$ KB\\
    \hline
  \end{tabular}
\vspace{-1em}
  \label{tab:benchmarks}
\end{table}

\subsection{Breakdown of End-to-End Inference Time}
\label{sect:time_breakdown}

\begin{figure}[t!] \centering
\includegraphics[width=0.495\textwidth]{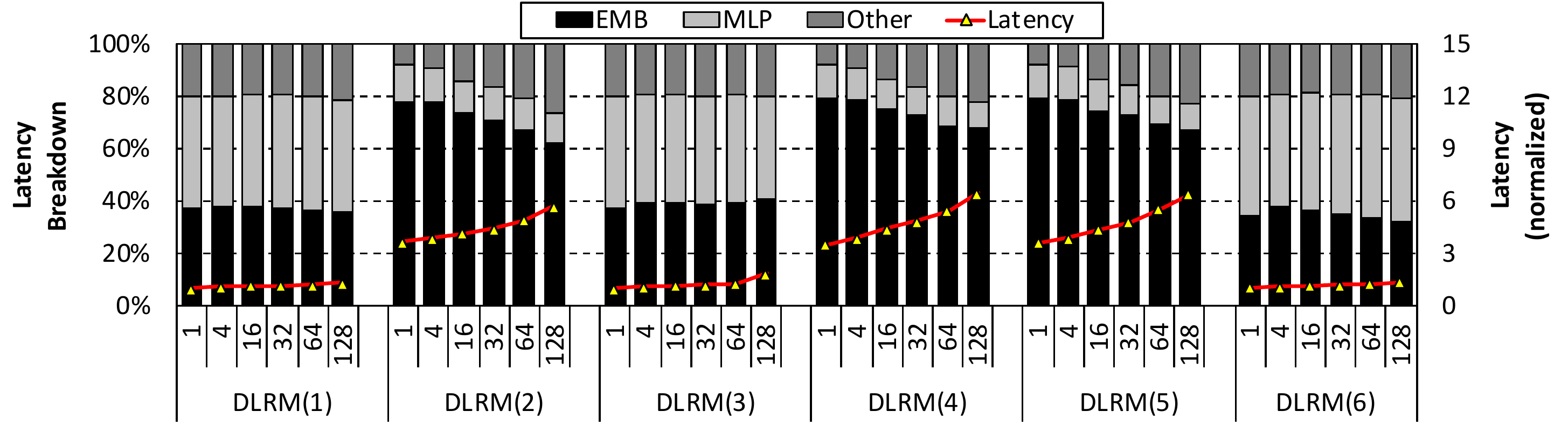}
\vspace{-1em}
\caption{
 Breakdown of CPU's inference latency into embedding layers (EMB), MLP layers,
					and others (left-axis) as a function of batch size, from $1$ to $128$ (x-axis). The
inference latency normalized to the slowest DLRM model with batch size $1$ (\dlrm{1})
is shown on the right-axis.
}
\vspace{-0.75em}
\label{fig:workload_latency}
\end{figure}

\fig{fig:workload_latency} shows a breakdown of end-to-end inference latency
and normalized execution time  when sweeping the input batch size  from $1$ to
$128$.  There are several interesting observations to be made from this
experiment.  First, unlike conventional ML applications extensively studied in
the computer systems community, \emph{non-DNN} layers such as embedding layers
take up significant fraction of execution time on personalized recommendation
models.  Second, MLP layers still account for a
non-trivial portion of runtime, especially when the inference batch size is
small.  Third, 	although larger batch sizes increase the latency of both the
embedding and MLP layers, MLP layers experience a relatively slower
increase in execution time than embedding layers (except for \dlrm{6} which is intentionally
	configured to have a lightweight embedding layer followed by a much more compute-intensive MLP layer, see \sect{sect:methodology}
	for details of our methodology).  This is
because large batch sizes tend to help increase the data reuse of MLP weights
across the multiple input batches and amortize the cost of uploading weights on-chip (e.g.,
as detailed in the next subsection, LLC miss rates in MLP layers are never more than $20\%$), 
			 whereas larger batches in embeddings do not
translate into better data reuse whatsoever. In other words, large batch sizes simply
result in a larger amount of embeddings to be gathered (\fig{fig:sls_code}) from the memory
subsystem which, depending on the relative execution time of embedding layers with respect to other layers, can
result in a proportional increase in execution time. In general, we conclude
that DNN-based recommendation systems are severely bottlenecked by embedding
layers. Nonetheless, MLP layers also account for a significant portion of
execution time, especially when the batch size is small for some configurations.

\subsection{On-chip Caching Efficiency}
\label{sect:caching}

\begin{figure}[t!] \centering
\subfloat[]
{
  \includegraphics[width=0.485\textwidth]{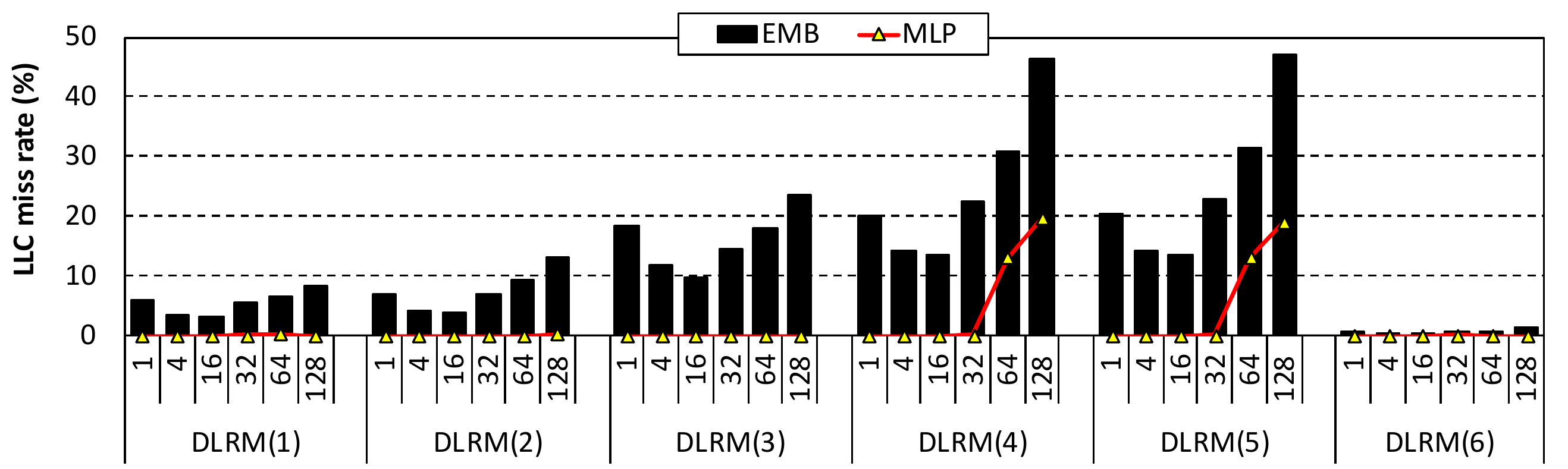}
}
\vspace{-0.3em}
\subfloat[]
{
  \includegraphics[width=0.485\textwidth]{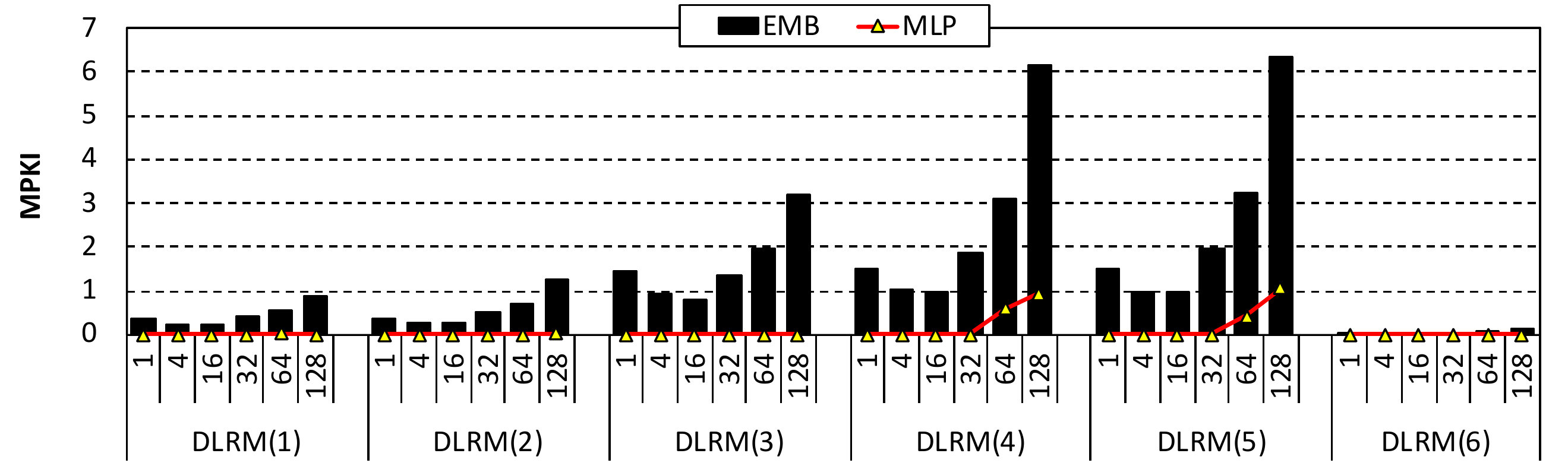}
}
\vspace{0.1em}
\caption{
Effect of executing embedding (EMB) and MLP layers on  (a) LLC miss rate and (b) MPKI
as a function of batch size (from $1$ to $128$). 
	We use Callgrind~\cite{valgrind} to collect the
	profiled statistics used for these experiments.
}
\vspace{-0.75em}
\label{fig:cpu_caching}
\end{figure}

To better understand the compute and memory bandwidth  demands of the
aforementioned two bottleneck layers (i.e., sparse embedding layers and MLP
		layers), we conduct a detailed analysis on the CPU's LLC miss rate and MPKI
(misses per thousand instructions) while executing embedding and MLP layers
(\fig{fig:cpu_caching}). In general, embedding layer's LLC miss rate shows high
sensitivity to input batch size with an increasing number of LLC
misses as batch size is increased.  The reason behind embedding layer's high
LLC miss rate is as follows.  A unique property of embedding tables is that its
size can be in the order of several tens to hundreds of
GBs~\cite{park:2018:fb,tensordimm,dlrm:arch}. This is because the total number
of embedding vectors within a table increases proportional to the number of
users/items (e.g., total number of users registered or movies serviceable in
		YouTube/Netflix).  As such, the embedding gather operations over such
high-capacity embedding tables are extremely \emph{sparse} with little
spatial/temporal locality. Now, the aggregate size of the gathered embeddings
scales up proportional to the batch size (\fig{fig:sls_code}), which directly
translates into higher memory traffic -- but one with low locality.  Larger
batch sized embedding layers therefore end up more severely pressurizing the
LLC, leading to larger number of LLC misses and  higher MPKI
(\fig{fig:cpu_caching}).

In terms of the MLP layers, the
LLC miss rate of these layers exhibit relatively less sensitivity to the input batch size because the
aggregate model size of the MLP layers in all our workloads are sufficiently
small enough (typically less than 1MB) to be captured inside the tens of
MBs of CPU on-chip caches. Therefore, the MLP layers in recommendation
models typically exhibit low LLC miss rates ($<$$20\%$) and low MPKI, exhibiting a
compute-limited behavior.

\subsection{Effective Memory Throughput}
\label{sect:mem_bw}

\begin{figure}[t!] \centering
\subfloat[]
{
  \includegraphics[width=0.485\textwidth]{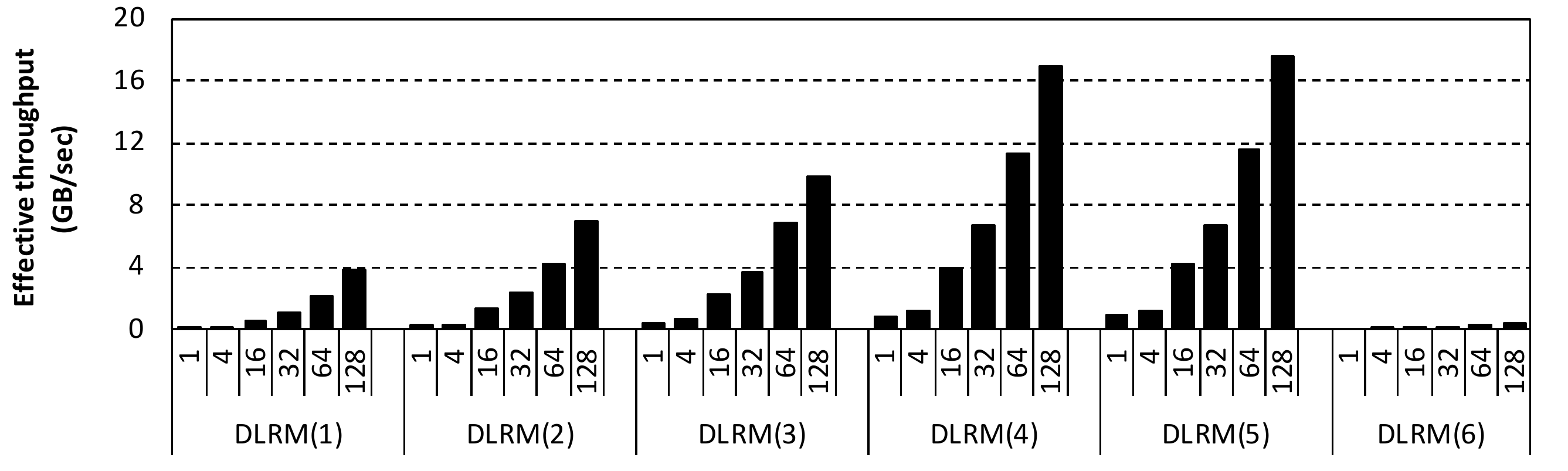}
}
\vspace{-0.4em}
\subfloat[]
{
\includegraphics[width=0.485\textwidth]{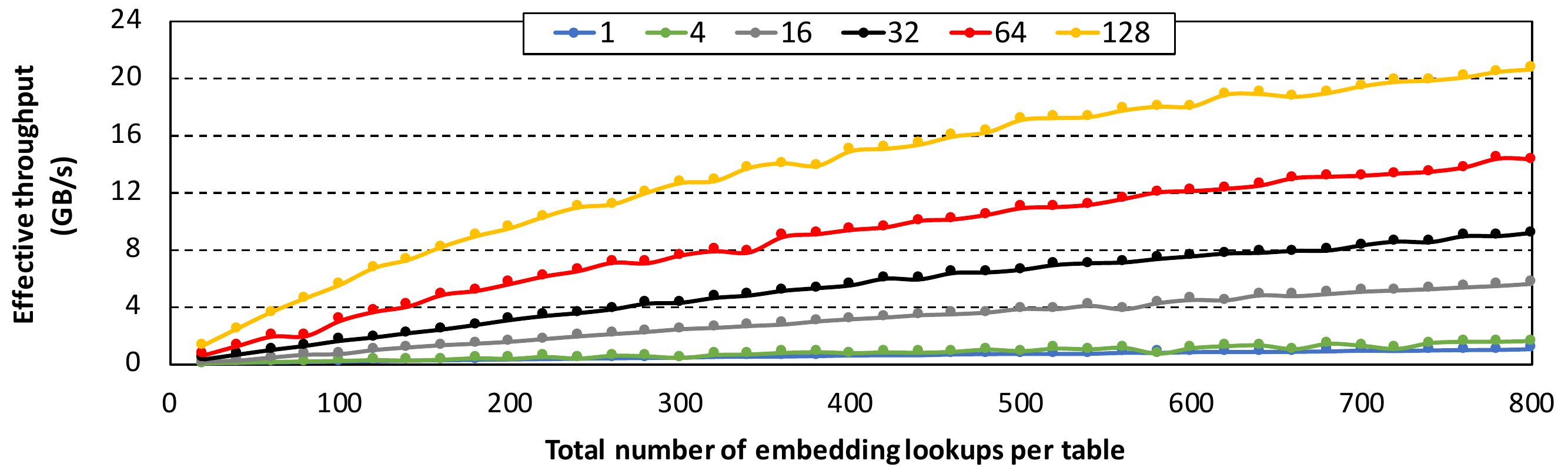}

}
\vspace{0.1em}
\caption{
	(a) Embedding layer's effective memory throughput for embedding
		gathers and reductions as a function of input batch size (from $1$ to
				$128$). To quantify its sensitivity to the number
		of embeddings gathered, the effective throughput of 
		a single table configuration in \dlrm{4} is plotted in (b) 
		when sweeping the total number of embeddings gathered. As depicted, the effective
		memory throughput generally grows monotonically as the batch
		size increases or when the number of embeddings gathered are increased. However, 
		the effective throughput is far below the maximum memory bandwidth, especially
		with small batch sizes or under realistic number of gathers per table (i.e., 
				typically under $100$	gathers per table~\cite{youtube_recsys,fox,facebook_dlrm,recnmp,tensordimm,deeprecsys}).
}
\vspace{-0.75em}
\label{fig:cpu_dram_bw}
\end{figure}

While sparse embedding layers exhibit a  high LLC miss rate and an
accordingly high MPKI (compared to MLP layers), we observe that the
\emph{``effective'' memory bandwidth utilized in gathering embedding vectors} 
is extremely low.  \fig{fig:cpu_dram_bw} summarizes the
	memory throughput of gathering embedding vectors while
	executing embedding layers.  To clearly quantify how efficiently memory
	bandwidth is being utilized for embedding lookups, we measure the
	\emph{effective memory throughput for embedding layers} by only considering
	the useful number of bytes transferred in gathering and reducing embeddings
	(i.e., size of total embedding vectors gathered / latency incurred in
			executing the embedding layer)\footnote{Directly measuring DRAM bandwidth utilization 
		using Intel VTune~\cite{intel:vtune} followed similar trends, albeit
		with smaller numbers than our defined effective memory throughput as subset of
		gathered embeddings can hit in the cache.
	}.  As depicted
	in \fig{fig:cpu_dram_bw}, the effective memory throughput for embedding
	layers is  far below the maximum $77$ GB/sec of memory bandwidth 
	of our baseline CPU memory system (\sect{sect:methodology}).
	Recall that a single embedding vector
	is only in the order of several hundreds of bytes (i.e., $128$ bytes
			with our default $32$-dimensional vector), far below the size
	of an $8$ KB of DRAM row buffer. 	Additionally, each of these vector loads
	have limited spatial locality due to their sparse and
	irregular memory access nature. 
Unlike throughput-optimized GPUs which
	execute with several thousands of concurrent threads with a large number of
	MSHRs (e.g., NVIDIA Volta's L1 cache implements the so-called \emph{streaming
			cache} which allows unlimited inflight cache misses to maximize data
			fetch throughput~\cite{volta:2017:hotchips}), latency-optimized CPUs
	utilize only tens of threads with a handful of MSHRs. As 
	the aggregate size of the
	embedding vectors gathered is only in the order of several KBs (low batch) or MBs (large batch)
	over several tens of GBs of embedding tables,
	it makes it challenging for CPU
	architectures to maximize memory-level parallelism and
	thus memory bandwidth utilization under the sparse, irregular, and fine-grained 
	vector gather operations\footnote{It is possible to achieve more than $50$ GB/sec of
		effective throughput ($>$$70\%$ of max) in embedding layers when the batch size is larger
			than $2048$ or when the embedding vector dimension is sufficiently large
			(i.e., more than $1024$-dimensional vector). However, such large batch size and wide embedding dimensions is an unrealistic one to assume for inference. 
	}.

\section{Centaur: A Hybrid Sparse-Dense Accelerator for Personalized Recommendation}
\label{sect:proposed}

We present \proposed, a chiplet-based hybrid sparse-dense accelerator that holistically
addresses the dual challenges of memory limited embeddings and compute limited
MLPs of personalized recommendations. 
To the best of our knowledge, \proposed is the first end-to-end accelerator that tackles
both the memory and compute bottlenecks of personalized recommendation models.
We first present our motivation for a
package-integrated CPU+FPGA platform (rather than ASICs),
	followed by a description of our proposed architecture.

\subsection{Motivation: Why Package-integrated CPU+FPGAs?}
\label{sect:motivation}

GPUs are currently the dominating processor architecture for ML training because their
throughput-optimized design suits well for
the (throughput-heavy) algorithmic nature of training.  For cloud deployment
of recommendation services however,
					latency-optimized CPUs are the preferred architecture of choice.
First, the abundance of readily available CPUs in today's datacenters makes it an appealing computing
platform from a total cost of ownership (TCO) perspective, especially when
considering the off-peak portions of the diurnal cycle where CPUs would
otherwise remain idle~\cite{hazelwood:2018:hpca}.  Second, user-facing
inference services (e.g., news feed, advertisement, e-commerce) for
recommendations have firm SLA (service level agreement) goals to meet 
which renders latency-optimized CPUs more suitable than throughput-oriented
GPUs. Lastly, recall that sparse embedding layers are significantly memory capacity hungry
because the embedding tables can require up to several hundreds of GBs of memory usage (\sect{sect:mem_bw}). As a result, the
bandwidth-optimized 3D stacked memory employed in GPUs or ML accelerators such
as Google TPUs~\cite{tpu1,tpu2} cannot store the embedding tables locally inside their physical memory,
	 preventing them from being used for deploying recommendations.  Therefore, 
	 the vast majority of cloud ML inference services for personalized
	 recommendation are primarily powered using \cpuonly systems as noted by several hyperscalers~\cite{dlrm:arch, park:2018:fb, hazelwood:2018:hpca}.

	 Given this landscape, we observe that package-integrated CPU+FPGAs become a promising
	 solution as it holistically addresses \emph{all} the aforementioned challenges, as detailed
	 below:

	 \begin{enumerate}
	 \item Package-integrated CPU+FPGAs are minimally intrusive to existing server chassis designs
	 (and therefore the server rack and the overall datacenter)
	 as they are \emph{socket} compatible to existing system nodes. Furthermore, CPUs can still function
	 as a ``\emph{host}'' from the OS's perspective (unlike GPUs/TPUs which are slave devices). As such, they can 
	 be utilized for \emph{non}-ML usages thus enhancing the resource utility for optimizing TCO.

	 \item 	 The reconfigurable FPGA logic can be utilized to address performance
	 bottlenecks, further reducing inference latency to help satisfy SLA goals and improve QoS.

	 \item More importantly, the CPU and FPGA both share a single physical memory (i.e., the memory DIMMs
		within/across CPU sockets) which is based on capacity-optimized DDRx. This allows the FPGA-side accelerator
	 to keep the entire embedding tables in CPU memory as-is and access them directly using the high-bandwidth,
	 low-latency CPU$\leftrightarrow$FPGA communication channels, 
		 a requirement \emph{discrete} GPUs or FPGAs cannot fulfill.
	 \end{enumerate}

	Based on these {\bf key observations}, our \proposed architecture utilizes
	the FPGA's programmable logic area to implement a \emph{heterogeneous}
	computing device, synergistically combining a sparse accelerator for
	embedding gathers/reductions and a dense accelerator for GEMM computations.
	Before we detail the microarchitecture of our sparse-dense accelerator, the
	next subsection first discusses our proposed chiplet-based CPU+FPGA architecture.
	We then discuss our proof-of-concept CPU+FPGA substrate, Intel HARPv2~\cite{intel:harpv2}, which we
	utilize to demonstrate our proposal.

\begin{figure}[t!] \centering
\includegraphics[width=0.23\textwidth]{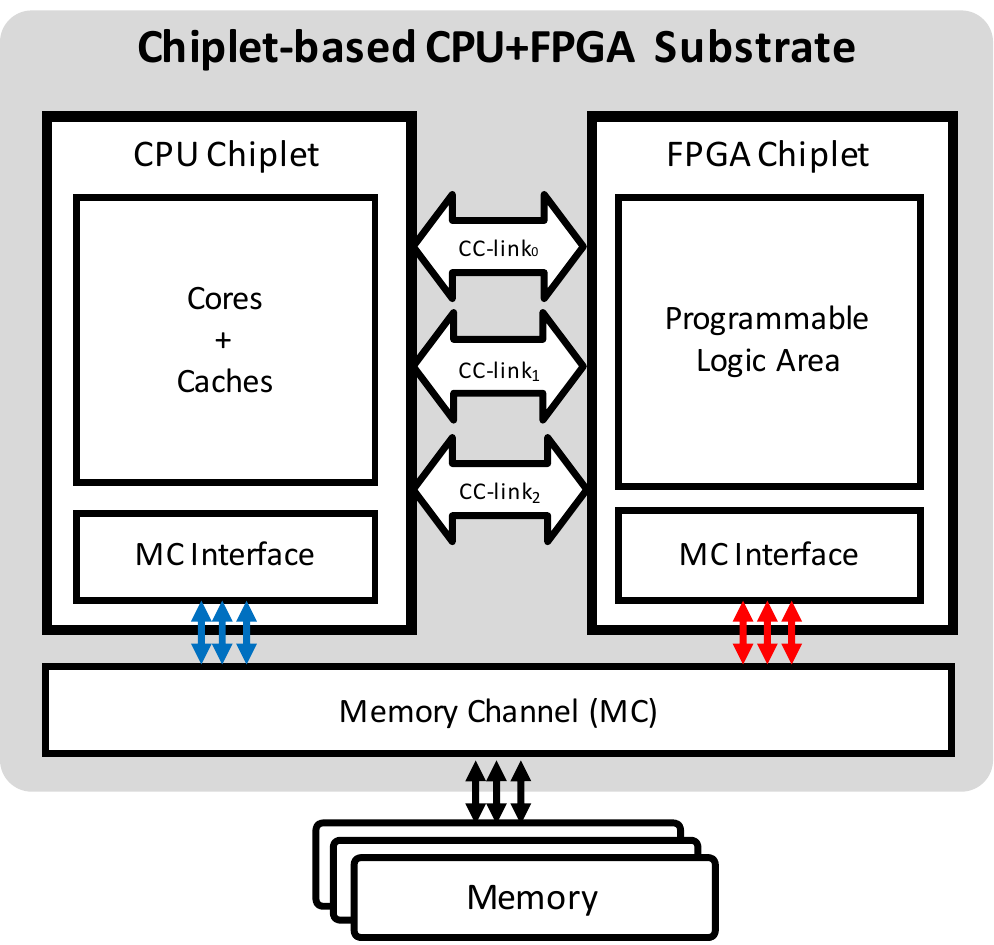}
\vspace{-0.5em}
\caption{
Proposed package-integrated CPU+FPGA architecture.
}
\vspace{-0.75em}
\label{fig:centaur_vision}
\end{figure}

\begin{figure*}[t!] 
\centering
\includegraphics[width=0.995\textwidth]{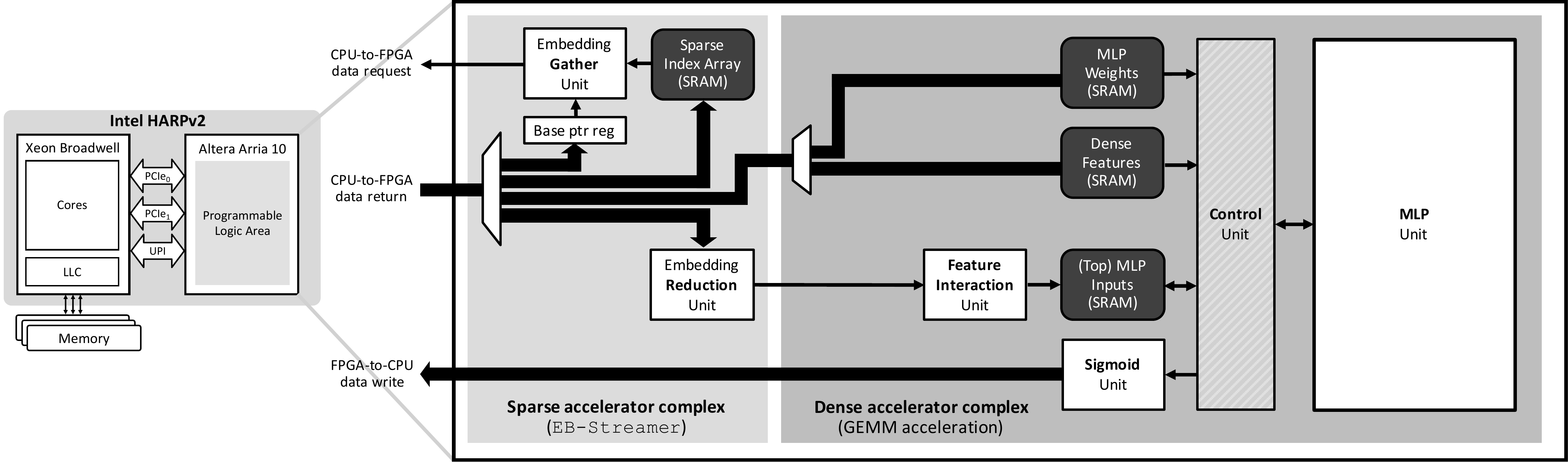}
\vspace{-0.5em}
\caption{
	High-level overview of our proposed \proposed architecture. As a proof-of-concept
		prototype, we	utilize Intel HARPv2 to design our
		hybrid sparse-dense accelerator. The reconfigurable FPGA logic
		are used to synthesize both the sparse (\streamer used for high-throughput,
				low-latency embedding gathers and reductions) and dense (for high-throughput GEMM computation) accelerators.
}
\vspace{-0.75em}
\label{fig:centaur}
\end{figure*}

\subsection{Proposed Chiplet-based CPU+FPGA Architecture}
\label{sect:vision}

\fig{fig:centaur_vision} illustrates our proposed chiplet-based CPU+FPGA
architecture, which is designed to be minimally intrusive to existing
TCO-optimized server chassis/rack as it is socket-compatible to current
systems.  The FPGA chiplet has two communication paths to the CPU memory
subsystem. The \emph{cache coherent path} utilizes the CPU$\leftrightarrow$FPGA
cache coherent links (denoted as CC-link$_{n}$) to traverse through the CPU
on-chip cache hierarchy first and then to the off-chip memory (via the
		blue-colored arrows), which can be effective for memory accesses with high
data locality. An alternative \emph{cache bypassing path} (via the red-colored
		arrows), which is more appropriate for our memory-intensive embedding layers, utilizes
a separate memory channel interface that completely \emph{bypasses} the CPU
caches and directly routes FPGA-side memory requests to the off-chip memory
interface. By provisioning the cache bypassing route's communication
throughput to be commensurate to (or higher than) the maximum off-chip memory bandwidth, the
sparse accelerator of \proposed can significantly boost the throughput of
embedding layers by conducting vector gathers over this communication channel.

Unfortunately, chiplet-based commercialized CPU+FPGA designs are still at an
early stage with limited accessibility and functionality. We therefore utilize
Intel's HARPv2~\cite{intel:harpv2} as a proof-of-concept substrate to demonstrate the
merits of our proposal. As we detail in the next subsection, HARPv2 comes with
the cache coherent path (but no cache bypassing route) for CPU memory accesses,
		so the throughput benefits of our sparse accelerator is constrained by the
		memory-level parallelism that can be reaped out over the CPU$\leftrightarrow$FPGA
		cache coherent path, and accordingly the CPU cache
		hierarchy. Nonetheless, we use it to conservatively estimate the
		throughput benefits chiplet-based CPU+FPGAs can provide for recommendations.
		In the following subsections, we first present the details of our
		sparse-dense accelerator microarchitecture, followed by a description of its
		software interface to the overall system.

\subsection{Sparse Accelerator}
\label{sect:sparse_acc}

\begin{figure}[t!] \centering
\includegraphics[width=0.485\textwidth]{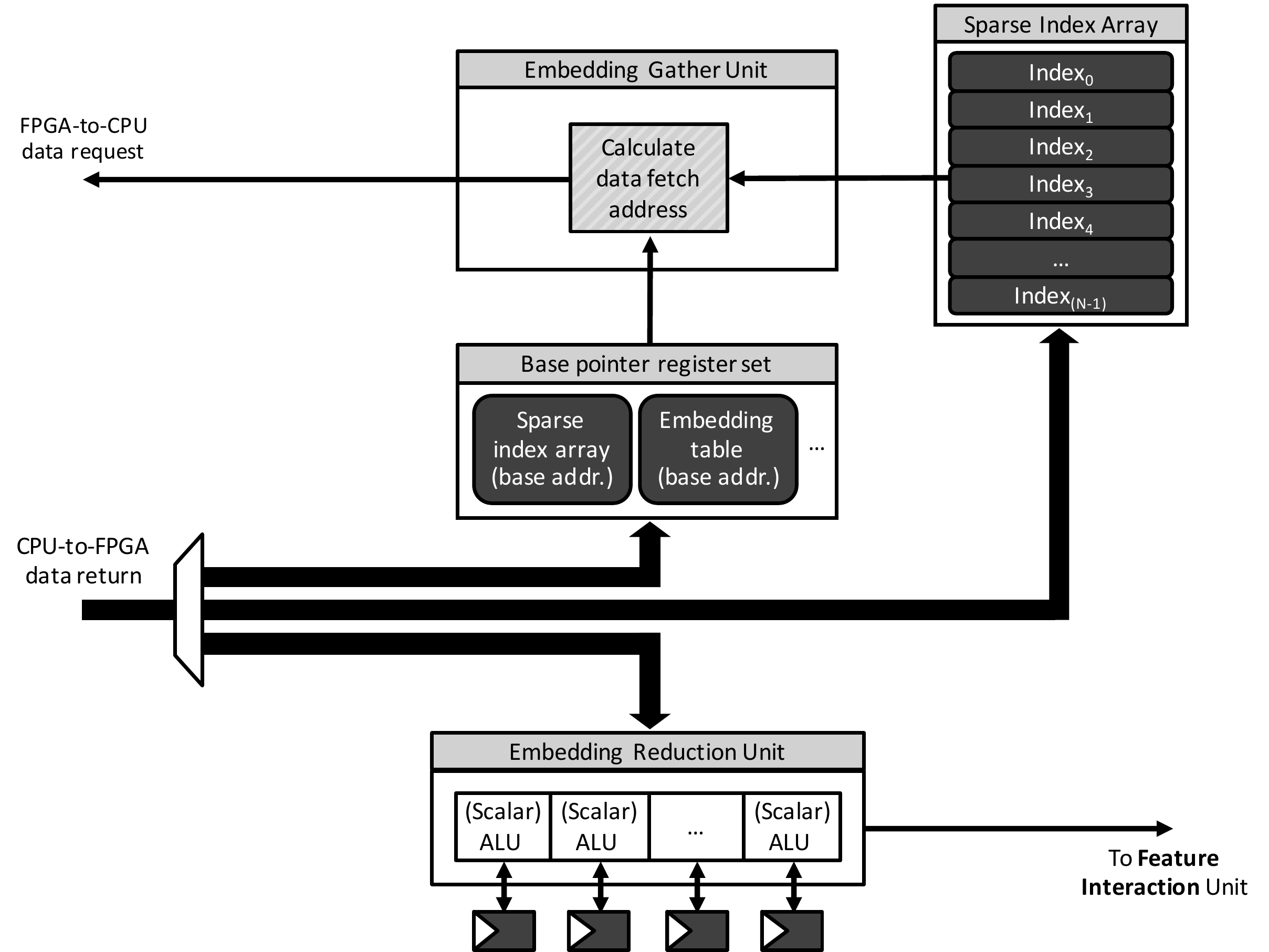}
\vspace{-0.7em}
\caption{
Microarchitecture of \proposed sparse accelerator.
}
\vspace{-0.75em}
\label{fig:uarch_sparse}
\end{figure}

 The key design objective of our sparse accelerator
is to enable high-throughput, low-latency embedding gather and reduction
operations.  Recall that package-integrated CPU+FPGA devices enable the
custom-designed FPGA logic to \emph{directly} access the shared physical
memory system in fine-grained ($64$-Byte) cache line granularity  via
cache-coherent high-bandwidth communication links. Under the Intel HARPv2
platform we assume in this work, a theoretical maximum uni-directional communication bandwidth
of $28.8$ GB/sec is provided between the CPU and FPGA using two PCIe links
and one cache coherent UPI link.
Our sparse
		accelerator utilize such communication technology to implement an
		\emph{embedding streaming unit} (henceforth referred to as \streamer) that
		spawns off multiple embedding vector gather operations followed by an on-the-fly reduction operation, in a high-throughput manner.
 \fig{fig:uarch_sparse} details the
		microarchitecture of \streamer, which contains a base pointer register set
		(\bpr), sparse index SRAM array (\spid), embedding gather unit (\ebgu), and the
		embedding reduction unit (\ebru). The embedding gathers and reductions are conducted 
		as follows:

\begin{enumerate}

\item When system is booted up, the CPU utilizes the MMIO interface to inform
the FPGA the CPU memory addresses that point to a) the sparse index array
(i.e., the row IDs to gather from the embedding table), b) the embedding table,
	c) the MLP weights, and d) dense features (to be used as inputs for the
	bottom MLP).  These base pointer values are copied into the \bpr to be
	utilized by the sparse-dense accelerators for both embedding gathers and GEMM
	operations.

\item Once \bpr is initialized, the \ebgu utilizes \bpr's base pointer address
of the sparse index array to perform a CPU$\rightarrow$FPGA read operation
which populates the \spid with sparse index IDs subject for gather operations. 
Notice that \ebgu is nothing more than an address
generator (i.e., base + offset, see \fig{fig:sls_code}) which is dominated by
logic gates, thus having low implementation overhead.

\item Using the embedding table base address value stored in \bpr and the
sparse index IDs stored in \spid, the \ebgu starts generating
CPU$\rightarrow$FPGA embedding gather operations. To maximally utilize
CPU$\leftrightarrow$FPGA communication bandwidth, the \ebgu monitors 
the communication bandwidth utility and aggressively instantiates embedding
vector read operations over the PCIe/UPI links, whenever the
CPU$\leftrightarrow$FPGA communication links become available.

\item When the embedding vectors arrive at the sparse accelerator, they
are immediately routed to our \ebru. As vector reductions are \emph{in-place}
operations, \ebru conducts embedding  ``reduction'' operations on-the-fly
whenever the embedding vectors are streamed into \ebru.

\item Once all embeddings are gathered and reduced, the \ebru forwards the
reduced embedding vector to the dense accelerator complex.

\end{enumerate}

As embeddings are typically sized as $32$-wide vectors, a single
embedding vector gather operation is equivalent to a $32\times4$$=$$128$-Byte
load instruction.  Note that the memory addresses of the multiple embedding vectors 
subject for gathering are scattered across the memory address space.  
Consequently, a brute-force, software level
			 data transfer over such fine-grained, irregular data access
			 stream can incur severe latency overheads as each
			 \texttt{cpuToFpgaMemcpy()} API execution for a $128$-Byte CPU$\rightarrow$FPGA 
			 read
			 operation must traverse through various layers in the software stack.
			 One of the key advantage of initiating embedding gather operations over
			 the package-integrated CPU$\leftrightarrow$FPGA channels is that the process of data fetch
			 and retrieval is entirely orchestrated at the hardware level,
			 significantly reducing memory access latency.  Furthermore, embedding
			 gathers are conducted while being less interfered and bottlenecked by the
			 CPU's cache hierarchy. As discussed in \sect{sect:characterization},
			 embedding gather operations are inherently sparse with extremely low
			 locality, rendering conventional CPU caching mechanism ineffective.
			 Nonetheless, the baseline \cpuonly system must always traverse through
			 the multi-level on-chip caches for all embedding vector load operations,
			 only to discover that the embeddings to be gathered are (most likely)
	located in CPU memory. Because the entire embedding gathering process is
	orchestrated using a handful of threads, \cpuonly embedding gathers are
	limited in terms of both parallelism and locality, achieving low memory
	bandwidth utility (\fig{fig:cpu_dram_bw}).  Because our sparse accelerator directly fetches
	the embeddings over the CPU$\leftrightarrow$FPGA communication links, \proposed
	can achieve significantly higher memory bandwidth utilization (\sect{sect:eval_bw}) and fundamentally address the memory
	bandwidth challenges of embedding layers.

\subsection{Dense Accelerator}
\label{sect:dense_acc}

\begin{figure}[t!] \centering
\includegraphics[width=0.485\textwidth]{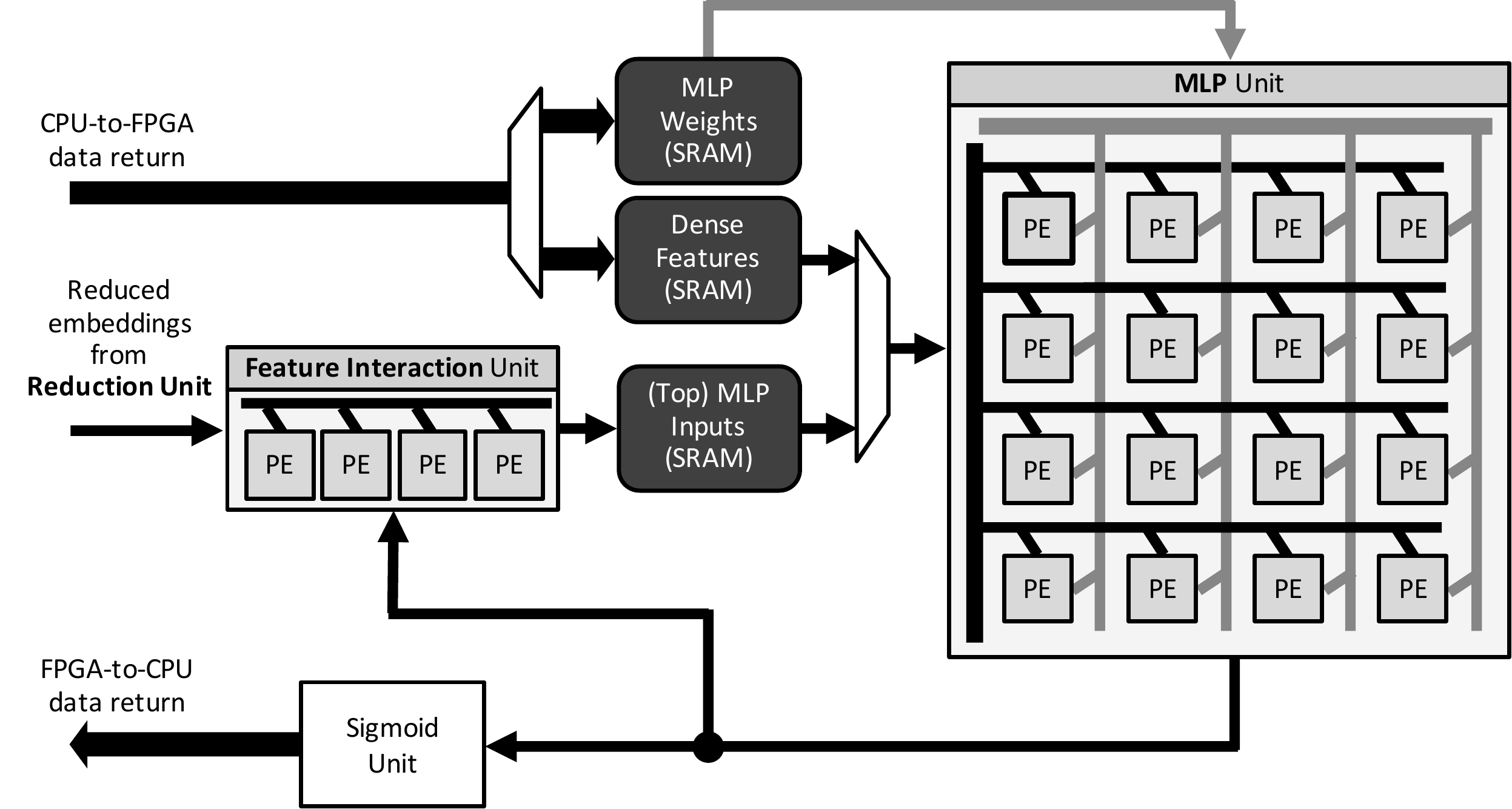}
\vspace{-0.7em}
\caption{
Microarchitecture of \proposed dense accelerator.
}
\vspace{-0.8em}
\label{fig:uarch_dense}
\end{figure}

\begin{figure}[t!] \centering
\includegraphics[width=0.37\textwidth]{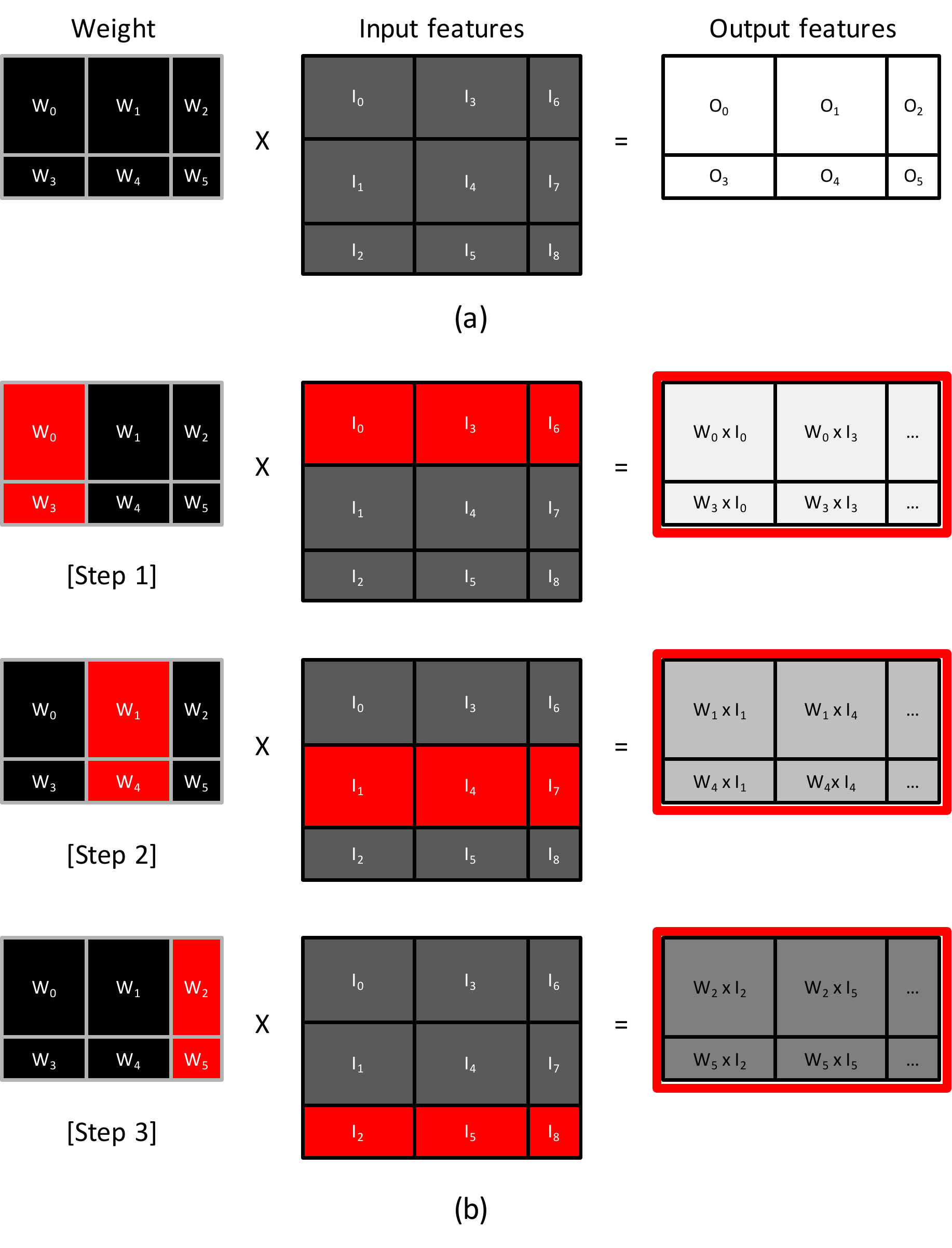}
\vspace{-0.5em}
\caption{
The output-stationary dataflow in \proposed's MLP unit's (a) GEMM operation. 
(b) An outer-product between the weight-input tiles generates the output tiles
to be accumulated into the intra-PE SRAM buffers. Each PE conducts 
a W$_{m}$$\times$I$_{n}$ matrix multiplication operation between the weight
and input tiles. In each computation step, a given W$_{m}$ tile (I$_{n}$ tile) is broadcasted to all the PEs
within its corresponding row (column) using the bus interconnection network within the MLP unit (\fig{fig:uarch_dense}).
}
\vspace{-0.75em}
\label{fig:gemm_dataflow}
\end{figure}

We now present our dense accelerator design, the microarchitecture of which is
shown in \fig{fig:uarch_dense}.  The primary design objective of our dense
accelerator is to speed up the execution of GEMM, the key algorithm that powers
both the MLP layers and the batched GEMM operation for feature interactions.
We use Altera's FPGA floating-point IP core~\cite{altera_fp_ip_core_manual} 
optimized for matrix multiplications
between two square matrices (the FP\_MATRIX\_MULT module) as key
building blocks to construct our dense accelerator complex. A processing engine
(PE) in \fig{fig:uarch_dense} is based on a single instance of the
 FP\_MATRIX\_MULT module (configured to handle matrix multiplication between two [$32\times32$] matrices),
which we utilize to compose a $4\times4$ spatial PE
array for the MLP unit and another four instances of PEs for the feature
interactions.  Putting all these together, \proposed provides an aggregate computational
throughput of $313$ GFLOPS operating over $200$ MHz.
The MLP control unit employs an output-stationary
dataflow~\cite{eyeriss_isca} which tiles the input and weight matrices in
[$32\times32$] sizes (to be compatible with the PE's GEMM compute
		granularity) and broadcasts these tiles across the spatial PE array.  The
MLP unit then conducts an outer-product among the input and weight tiles using
the PE array, which generates the partial sums to be temporally accumulated
into the SRAM buffers allocated per each PE (\fig{fig:gemm_dataflow}).  In
addition to the GEMM computation units, the dense accelerator complex contains
several SRAM buffers to store 1) the MLP weights (\mlpmodel), 2) the dense
features to be used as inputs to the bottom MLP layers (\densefeature), and 3)
	the (top) MLP inputs (\mlpinput).  The model parameters that are used to
	execute both top and bottom MLP layers are copied over the CPU$\leftrightarrow$FPGA
	communication link using the \bpr at boot-time. The MLP weight values remain
	persistent throughout the entire deployment process, so the overhead of
	uploading model weights to the FPGA's \mlpmodel is negligible as it is
	amortized over all future inference requests serviced by \proposed. Using these
	modules, the dense accelerator complex goes through the following steps to
	finalize the recommendation process.

\begin{enumerate}
\item The \bpr in the sparse accelerator complex is used to upload the MLP weights into \mlpmodel and
the inputs to the bottom MLP layer into \densefeature. As noted above, initializing the \mlpmodel with model
parameters only has to be done once as they remain persistent, whereas \densefeature needs to be updated 
whenever there is a new inference request.

\item The MLP unit first uses \mlpmodel and \densefeature to execute the bottom MLP layer, the result of which
is forwarded to the feature interaction unit. 

\item Once the sparse accelerator forwards the reduced embeddings to the
feature interaction unit, the output vector of the bottom MLP
layer is concatenated with the reduced embeddings to form a tensor. The feature interaction
unit utilizes the concatenated tensor to 
initiate a batched GEMM computation for feature interactions (\fig{fig:sls_example}),
				 the result of which is stored into \mlpinput.
					
\item The outputs of the feature interaction unit, which is read out of \mlpinput, is subsequently routed to the MLP unit 
to execute the top MLP layers using the model parameters stored inside \mlpmodel.

\item Once the top MLP layers complete execution, the final results are forwarded to the Sigmoid unit to calculate the event
probability. The final result is then copied back to the CPU memory for post-processing.

\end{enumerate}

As the entire dense GEMM computation is orchestrated seamlessly with the sparse accelerator, \proposed provides
significantly higher throughput and reduced latency in executing dense DNN layers compared to \cpuonly systems.
In the following
subsection, we detail the software interface that enables CPU+FPGA integration into the overall system.

\subsection{Software Interface}
\label{sect:sw_interface}

As the package-integrated HARPv2 platform
provides a unified virtual memory address space between the CPU and FPGA,
				 the CPU+FPGA functions as a
single processor as far as the operating system and its applications are
concerned, supporting the ``pointer-is-a-pointer'' like semantics.  Concretely,
	the pointers to the sparse index array, the embedding tables, the dense
	feature inputs, and others are forwarded to the FPGA using the MMIO
	interface. As these base address pointers are virtual addresses, the
	FPGA-side IOMMU (and TLB) translates them into physical addresses when the
	embedding gather operations are conducted, allowing the FPGA to directly
	access the CPU physical memory at the hardware level. Compared to invoking
	multiple software invoked DMA copy operations, such fine-grained hardware
	level data movement helps reduce average memory access latency, allowing
	\proposed to achieve superior memory throughput for embedding
	gathers.  Once the base pointer address values for key data structures (e.g.,
			sparse index array, embedding tables, $\ldots$) are copied over to the
	\proposed's \bpr over MMIO, the inference process  is entirely orchestrated
	under the hood at the hardware level.  As a result, high-level ML framework
	(e.g., TensorFlow, PyTorch) can readily employ our proposed architectural
	solution with minimal changes.

\section{Methodology}
\label{sect:methodology}

{\bf Evaluation platform.}		We demonstrate and benchmark \proposed
on Intel HARPv2 system containing a Broadwell Xeon
E5-2680v4 and Altera Arria 10 GX1150~\cite{intel:harpv2}.  At the time of this
writing, Intel's HARPv2 platform (released in $2016$) is the \emph{only}
publicly accessible package-integrated x86 CPU+FPGA so we evaluate
\proposed using this computing architecture as a proof-of-concept prototype.
The	entire sparse-dense accelerator is written in SystemVerilog RTL and we use
		Quartus Prime Pro $16.0$ to synthesize, place, and route our design
		(\tab{tab:fpga}).  We explore three design points of recommender
		systems. The baseline \cpuonly uses HARPv2's Broadwell CPU without
		the FPGA activated for a fair comparison with \proposed.  Aside from
		\proposed, we also established an additional design point to better
		cover the design space of recommendation inference systems.  While
		CPUs are the preferred system design point in deploying 
	 recommendations (as discussed in
					\sect{sect:motivation}), we nonetheless evaluate the performance of a
				GPU-based system for the completeness of our study. Here, we assume the
				entire embedding tables are stored in CPU memory so once all the
				embedding vectors are gathered and reduced by the CPU (using
						\texttt{SparseLengthsSum()}, \fig{fig:sls_code}), the CPU copies
				them over PCIe to the GPU  for GPU-side MLP
				computation (referred to as \cpugpu~\cite{tensordimm}). We utilize
				NVIDIA DGX-1~\cite{dgx_1v} for \cpugpu performance measurements.
						When estimating CPU's power consumption, we used \texttt{pcm-power}
						for both CPU socket-level power estimation as well as the power
						consumed by its memory DIMMs. For GPU power consumption, NVIDIA's
						\texttt{nvprof} profiling tool has been utilized. For \proposed's
						CPU+FPGA power measurements, we use \texttt{pcm-power} to measure
						both the socket-level CPU+FPGA as well as the power consumed by the
						memory DIMMs.  When evaluating energy-efficiency, we multiply the
						power estimation values with each design-point's end-to-end
						inference execution time.  All performance numbers are measured end-to-end in wall clock time, 
						which is collected
after sufficiently warming up the CPU's cache hierarchy.

\begin{table}[t!]
  \centering
  \caption{Centaur FPGA resource utilization.}
\scriptsize
    \begin{tabular}{|c|ccccc|}
    \hline
	    & \textbf{ALM } & \textbf{Blk. Mem} & \textbf{RAM Blk.} & \textbf{DSP } & \textbf{PLL} \\
    \hline
    GX1150 (Max)			& 427,200		& 55.5 M	& 2,713 & 1,518	& 176\\
    \proposed					& 127,719		& 23.7 M	& 2,238 & 784		& 48 \\
    \hline
		\hline
		Utilization [\%]	& 29.9			& 42.6		& 82.5	& 51.6	& 27.3 \\ 
		\hline
  \end{tabular}
\vspace{-1.0em}
  \label{tab:fpga}
\end{table}

{\bf Benchmarks.} We use the
open-sourced deep learning recommendation model (DLRM) as our primary
benchmark suite~\cite{facebook_dlrm}.
DLRM is configured using the latest PyTorch backend library (version $1.5$ nightly
		build, accessed March $25$, $2020$) which extracts parallelism
	using OpenMP and AVX instructions
for embedding and MLP layers.
DLRM provides three reference
model architectures which are used across two different services and have
different configurations depending on their use-case. The configurations vary
in terms of the number of embedding tables, the number of gathers per each
embedding table, total memory requirement of embedding tables, and the number
of MLP layers and its dimension size. While maintaining the distinctive
characteristics of the default three models, we add three more configurations
to better highlight the different compute and memory access behavior of
recommendation models, as detailed in \sect{sect:characterization}.
\tab{tab:benchmarks} summarizes the six benchmarks we study in
this paper. Note that \dlrm{6}'s embedding layer has been artificially scaled down
to have a short embedding layer stage with a relatively
longer MLP computation step, which we utilize to evaluate \proposed's sensitivity to
MLP intensive recommendations.

\section {Evaluation} 
\label{sect:evaluation}

This section explores three design points of recommender systems: 1)
baseline \cpuonly, 2) \cpugpu, and 3) \proposed. We first discuss the FPGA resource utility of our
hybrid sparse-dense accelerator. We then compare the memory throughput and overall performance of
\cpuonly vs. \proposed, followed by a comprehensive comparison study between all three design 
points in terms of energy-efficiency.

\begin{table}[t!]
\caption{Sparse vs. Dense FPGA resource usage.}
\scriptsize
\vspace{-1.5em}
\begin{center}
\begin{tabular}{|c|c|cccc|}
\hline
	& \textbf{Module} & \textbf{LC comb.} & \textbf{LC reg.} & \textbf{Blk. Mem} & \textbf{DSP}\\ 
\hline
    \hline
    \multirow{5}{*}{Sparse}
		&Base ptr reg.						& 98		& 211		&	0			&0\\
		&Gather unit							& 295		& 216		& 0			& 0 \\
		&Reduction unit						& 108		& 8,260	&	0			& 96\\
		&SRAM arrays							& 350		& 98		&	12.2M & 0\\
		 & \textbf{Total}					&\textbf{851} & \textbf{8.8K} & \textbf{12.3M} & \textbf{96}\\
		\hline
    \hline
    \multirow{5}{*}{Dense}
		&MLP unit								& 40K 	& 131K	&	2.3M	& 512 \\
		&Feat. int. unit				& 10K		& 33K		&	593K	& 128	\\
		&SRAM arrays						& 1K		& 11K		&	1.6M	& 48\\
		&Weights								& 13		& 77		& 5.2M	& 0 \\
		 & \textbf{Total} & \textbf{52K} & \textbf{175K} & \textbf{9.8M} & \textbf{688} \\
		\hline
    \hline
		Others			& Misc. & 587 & 6K & 608K & 0 \\

    \hline
\end{tabular}
\end{center}
\label{tab:resrc_per_usage}
\vspace{-1em}
\end{table}

\subsection{Centaur FPGA Resource Utilization}
\label{sect:fpga_resrc}

\tab{tab:resrc_per_usage} summarizes how \proposed's sparse-dense accelerator
utilizes the various FPGA resources. As the major role of our sparse-optimized
accelerator is to perform high-throughput embedding gathers/reductions, the
\streamer is designed to incorporate a local sparse index array to be able to
seamlessly invoke multiple embedding gather operations in parallel. That is, we
employ a large SRAM array to hold many sparse index IDs such that the embedding
gather unit can aggressively launch multiple
gather operations concurrently, boosting memory-level parallelism and overall
memory bandwidth utilization.  This is reflected by the sparse accelerator
complex using $54\%$ of the block memory bits to store sparse indices,
				with little usage of the ALMs and DSPs ($6\%$ and $12\%$ usage,
						respectively) as the primary computation conducted inside the
				sparse accelerator is the address generation for gathers and
				reductions, both of which can be designed in a lightweight fashion.  The dense
				accelerator complex on the other hand is designed for high
				computational throughput, so it consumes $88\%$ of the DSPs and
					$94\%$ of the ALMs, achieving much higher computation throughput than
					\cpuonly systems. As we further discuss in the remainder of this section,
					such rather skewed, heterogeneous usage of FPGA resources helps
					\proposed strike a balance that effectively tackles the bottlenecks
					of memory intensive embedding gathers and compute limited
					GEMM operations.

\begin{figure}[t!] \centering
\subfloat[]
{
  \includegraphics[width=0.495\textwidth]{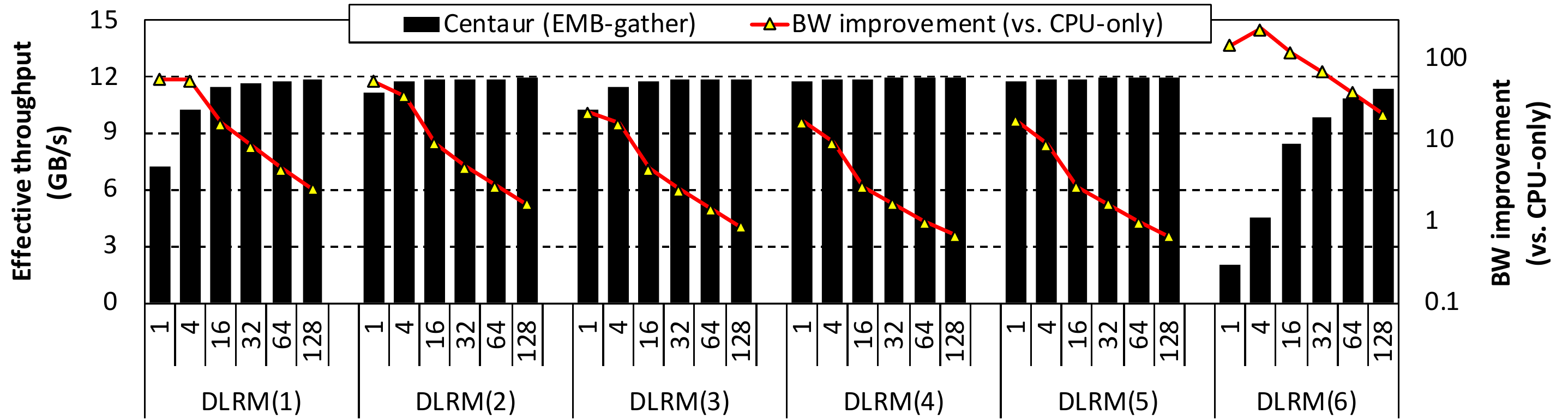}
}
\vspace{-0.4em}
\subfloat[]
{
\includegraphics[width=0.485\textwidth]{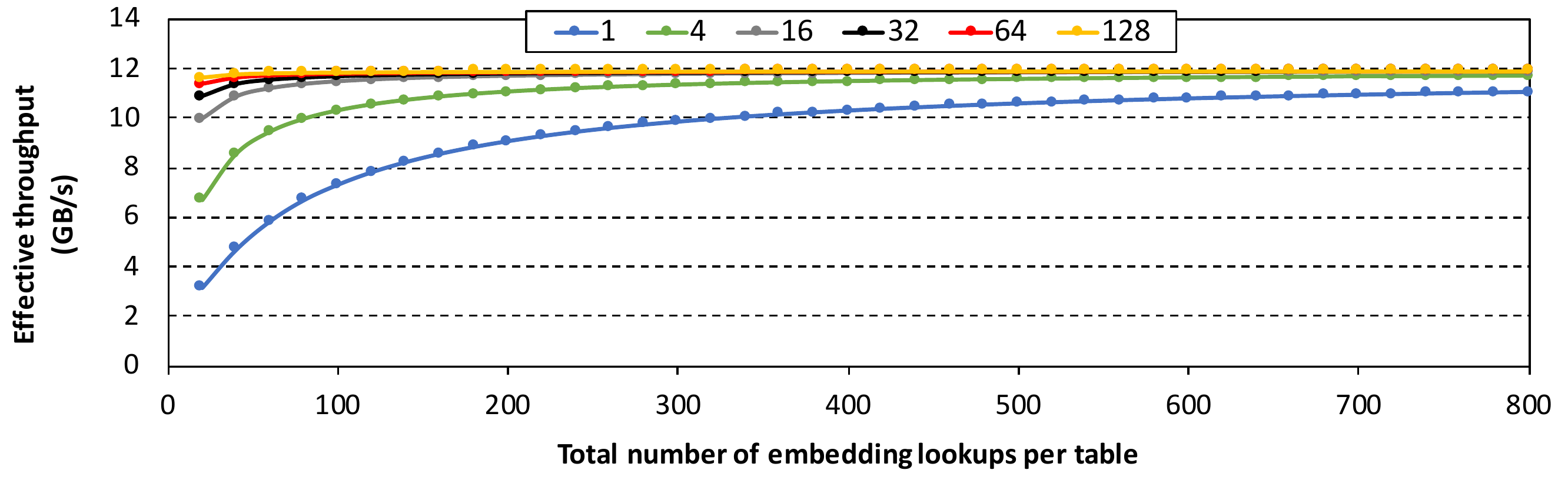}
}
\vspace{-0.4em}
\caption{
(a) \proposed's effective memory bandwidth utilized for embedding gathers (left-axis) 
	and its improvements compared to \cpuonly (right-axis)
	as a function of 
	input batch size (from $1$ to $128$).
	(b) \proposed's effective memory bandwidth as a function of 
	total number of embeddings gathered from the embedding tables,
	exhibiting a much rapid improvement in effective throughput than 
		the baseline \cpuonly (\fig{fig:cpu_dram_bw}(b)).
}
\vspace{-0.75em}
\label{fig:eval_bw}
\end{figure}

\subsection{Effective Memory Throughput for Embedding Layers}
\label{sect:eval_bw}

\cpuonly cannot effectively execute embedding layers because of its low memory
throughput in gathering embeddings, spending significant faction of time on
this bottleneck layer.  Our \streamer significantly
improves the effective throughput in gathering embedding
vectors, especially for low batches, achieving up to $11.9$ GB/sec of 
throughput (\fig{fig:eval_bw}). As the maximum possible \emph{effective}
uni-directional CPU$\leftrightarrow$FPGA communication bandwidth  is around
$17$$-$$18$ GB/sec in HARPv2, our \streamer achieves $68\%$ of the possible
communication bandwidth.  Given the highly irregular, sparse data access
patterns of embedding gathers, \streamer's high communication bandwidth utility
demonstrates the robustness of our embedding gather unit.  Because large
batches help \cpuonly better utilize memory bandwidth
(\fig{fig:cpu_dram_bw}(a)), the gap between \cpuonly and \proposed's memory
throughput gradually shrinks as batch size is increased. In particular,
					 \streamer falls short than \cpuonly by $33\%$ for \dlrm{4} and
					 \dlrm{5} with a large batch size of $128$, as \streamer's throughput
					 is constrained by the CPU$\leftrightarrow$FPGA link bandwidth.  As
					 detailed in \sect{sect:eval_perf}, such performance overhead for
					 large batches is offset by the high-throughput \proposed's dense
					 accelerator delivers.  Note that the effective
					 throughput of \streamer is expected to naturally scale up as
					 CPU$\leftrightarrow$FPGA communication link bandwidth is increased
					 with the latest high-bandwidth package-level signaling
					 technologies~\cite{intel:emib,intel:foveros,mcm_gpu,simba}.
					 Overall, \proposed provides an average $27\times$ throughput
					 improvement than \cpuonly across our studied configurations, even
					 with our conservatively chosen HARPv2 platform, thus effectively
					 tackling the memory bandwidth limitations of embedding layers. We
					 now discuss the end-to-end performance improvement our \proposed
					 delivers using our sparse-dense hybrid accelerator architecture.

\begin{figure}[t!] \centering
\includegraphics[width=0.495\textwidth]{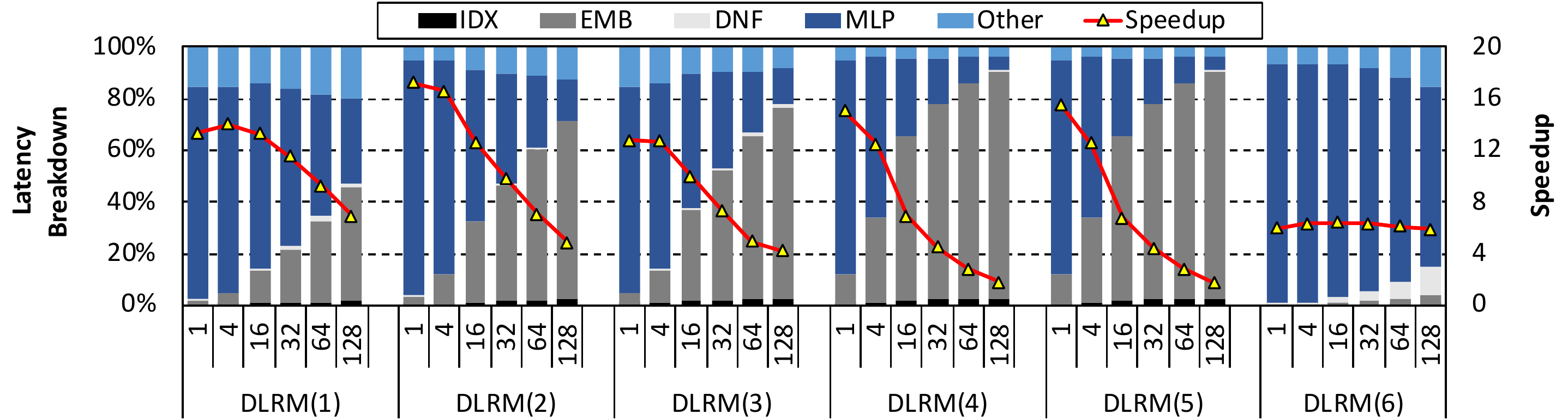}
\vspace{-1em}
\caption{
Breakdown of \proposed's inference time into CPU$\rightarrow$FPGA sparse index fetch time (IDX), embedding gathers/reductions (EMB),
 CPU$\rightarrow$FPGA dense feature fetch time (DNF), MLP execution, and others (left axis).
	 The right-axis summarizes the performance improvement \proposed	achieves compared to \cpuonly. 
}
\vspace{-0.75em}
\label{fig:eval_perf}
\end{figure}

\subsection{Performance} 
\label{sect:eval_perf}

\proposed significantly improves the performance of memory limited embedding
layers, thanks to \streamer's high-throughput gather operations. At the same time,
	the abundant computation units in dense accelerator complex reduces the latency to execute 
	GEMMs in recommendation models. This allows \proposed to substantially
	reduce end-to-end latency  as it holistically addresses the two most
	significant bottlenecks of recommendation.
	\fig{fig:eval_perf} shows a latency breakdown of our studied
workloads and the resulting performance improvement against baseline \cpuonly, achieving
$1.7$$-$$17.2\times$ end-to-end speedup. Among the six DLRM models we study, five of them
are bottlenecked by embedding layers especially under low batches, so the throughput-optimized \streamer helps
resolve the system bottlenecks, achieving superior performance improvements.
\dlrm{6} achieves a modest $6.2\times$ average speedup, which is expected
because this model is intentionally configured to have a heavyweight
MLP layer with a lightweight embedding layer (\tab{tab:benchmarks}). Consequently, the overall performance
is relatively insensitive to the improved memory throughput \streamer brings about. 
Nonetheless, \proposed's dense accelerator still provides significant latency reduction 
when executing \dlrm{6}'s GEMM, achieving substantial end-to-end performance improvement.

\subsection{Power and Energy-Efficiency}
\label{sect:eval_energy}

\begin{figure*}[t!] \centering
\subfloat[]
{
  \includegraphics[width=0.995\textwidth]{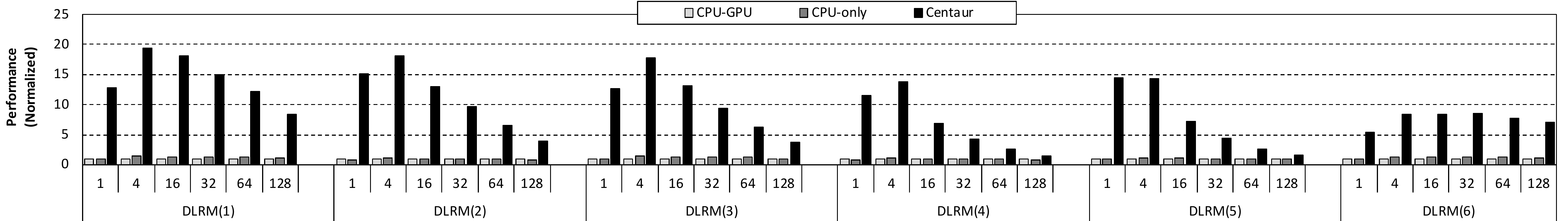}
}
\vspace{-0.3em}
\subfloat[]
{
  \includegraphics[width=0.995\textwidth]{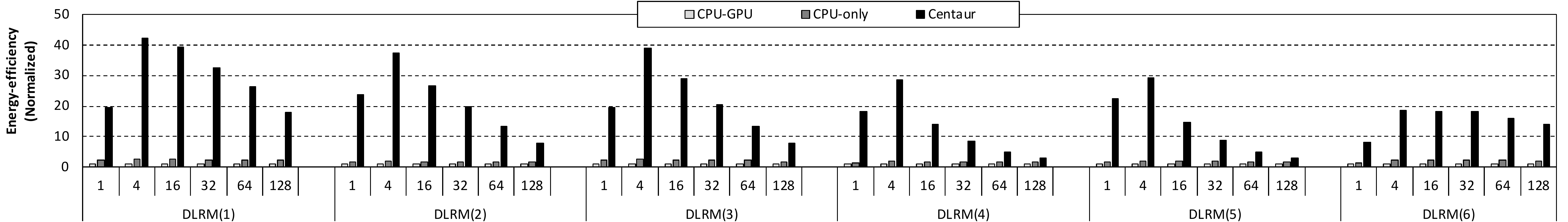}
}
\vspace{-0em}
\caption{
\proposed's (a) performance and (b) energy-efficiency improvement compared to \cpuonly and \cpugpu. All
	results are normalized to \cpugpu which exhibits the lowest performance and energy-efficiency.
}
\vspace{-0.75em}
\label{fig:comparison_all}
\end{figure*}

So far we have demonstrated the superior memory bandwidth utility and
performance of \proposed against the baseline \cpuonly. 
This section provides a comparison study of \proposed against \cpuonly and
\cpugpu in terms of power and energy-efficiency. \tab{tab:power} summarizes the
power consumption of our evaluated systems, the methodology of which is
summarized in \sect{sect:methodology}.  Compared to the baseline \cpuonly or
the power-hungry \cpugpu, \proposed consumes much less power, as the CPU cores
mostly remain idle while the FPGA-side sparse-dense accelerator orchestrates
the embedding gathers/reductions and the backend MLP computation step in a
power-efficient manner. As \proposed achieves superior power-efficiency while
also significantly improving end-to-end inference time, the overall
energy-efficiency is also significantly improved.  \fig{fig:comparison_all}
provides a summary of the performance and energy-efficiency improvements
\proposed brings about.  In general, the baseline \cpuonly performs better than
\cpugpu on average, achieving $1.1\times$ and $1.9\times$ performance and
energy-efficiency improvements. As the \cpugpu design needs to store the
embedding tables inside the CPU memory, the CPU-invoked embedding
gathers/reductions must always copy the reduced embeddings to the GPU for MLP
acceleration. This causes a noticeable latency penalty due to the
CPU$\rightarrow$GPU communication overhead, rendering \cpugpu to perform poorly
than \cpuonly. \proposed on the other hand achieves 1.7$-$17.2$\times$
performance speedup and 1.7$-$19.5$\times$ energy-efficiency improvement than
\cpuonly.  

\begin{table}[t!]
  \centering
  \caption{Power consumption.}
\scriptsize
    \begin{tabular}{|c|ccc|}
    \hline
	    & \textbf{CPU-only} & \textbf{CPU-GPU} & \textbf{Centaur}  \\
    \hline
    Power (Watts)  & $80$ & $91$/$56$ (CPU/GPU) & $74$ \\
    \hline
  \end{tabular}
\vspace{-0.75em}
  \label{tab:power}
\end{table}

\section{Discussion}
\label{sect:discussion}


Given the limited availability of chiplet-based, package integrated CPU+FPGA devices,
we utilized Intel HARPv2 as a proof-of-concept substrate to demonstrate the merits
of \proposed. With recent advances in packaging (e.g., Intel EMIB~\cite{intel:emib}
		and Foveros~\cite{intel:foveros}) and package-level signaling technologies (e.g.,
			NVIDIA's ground-referenced signaling~\cite{mcm_gpu,simba}),
architects are provided with rich a set of tools for designing chiplet-based
CPU+FPGA architectures. This section discusses some key design paramaters
of CPU+FPGAs and its implication in designing accelerators for recommendations. 

{\bf CPU$\leftrightarrow$FPGA bandwidth.} While our
	baseline CPU+FPGA platform provides \emph{only} $28.8$ GB/sec of
		CPU$\leftrightarrow$FPGA uni-directional communication bandwidth, upcoming
		package-level signaling technologies are expected to deliver several
		hundreds of GB/sec of communication throughput across
		chiplets~\cite{mcm_gpu,simba}. As discussed in
		\sect{sect:characterization}, the limited parallelism and throughput in
			gathering embedding vectors is one of the key obstacles for
				\texttt{CPU-only} designs. Note that embedding gather operations are
				inherently a collective operation where all embedding vectors must be gathered
				first in order to proceed to the following feature interaction stage.
				Because a significant fraction of vector reads are cache misses
				however, the gathering embeddings suffer from
				significant latency overheads due to the implicit barrier enforced in
				gathers.  An interesting CPU+FPGA design point is to optimize the
				overall architecture for throughput, rather than locality, and allow
				the high-bandwidth FPGA$\rightarrow$CPU embedding vector read
				operations to bypass the CPU cache hierarcy (as discussed in \sect{sect:vision}, \fig{fig:centaur_vision}),  maximizing
				available parallelism and throughput. Care must
				be taken however to guarantee cache coherence and consistency,
				which require carefully co-designed cache primitives for sparse
					embedding layers. Exploring such design
				point is part of our next future work.

{\bf FPGA size.} State-of-the-art FPGA-based dense accelerators
	provide several tera-operations scale of throughput, thanks to the abundant
		reconfigurable logic units available within the latest FPGA device (e.g.,
				Cloud-DNN provides $1.8$ TOPS of throughput over a Xilinx VU9P
				board~\cite{cloud_dnn}).  Given the embarrassingly parallel nature
			of DNN algorithms, we expect the effective throughput of our dense
				accelerator to proportionally scale up once the latest FPGA technology is
				integrated with the CPU. This 
				must of course be accompanied
				by a high-throughput CPU$\leftrightarrow$FPGA communication channel
				across chiplets in order to to proportionally feed enough
				input tensors to the accelerator, which can be delivered using
				the aforementioned, high-speed/high-bandwidth package-level signaling
				technology.

\section{Related Work}
\label{sect:related}

Recommendation models are the backbone ML algorithm that supports a variety of
internet services thus having significant industrial importance. While several
hyperscalers~\cite{dean:2018:goldenage,park:2018:fb,dlrm:arch,hazelwood:2018:hpca,
	hestness:2019:ppopp,dlrm:arch} hint at the scale of compute and memory
	required to deploy recommendations, little attention has been paid from the
	computer systems community to address this important research space (e.g., Wu
			et al.~\cite{sigarch:blog:recsys} states that only $2.1\%$ of research
			papers published in top computer architecture venues studies
			recommendation models).  
	Our recent work on TensorDIMM~\cite{tensordimm} was one of those few earlier
	works~\cite{tensordimm,recnmp,deeprecsys} in the architecture community
	to explore this research area, proposing  a hardware/software co-design for embedding layers.
	TensorDIMM employs a DIMM-based near-memory processing
	unit~\cite{chameleon,mcn} in a disaggregated GPU memory system as means to
	overcome the memory bandwidth bottlenecks of embedding layers. While the
	problem space Kwon et al. tackles is identical to \proposed, the following
	factors render our study unique compared to TensorDIMM. First, the focus of
	our work is on \emph{CPU-centric} systems which is the most commonly adopted
	inference deployment setting by hyperscalers, unlike TensorDIMM which assumes
	a \emph{GPU-centric} system for inference.  Second, TensorDIMM requires a
	separate, pooled memory architecture to achieve maximum performance benefits,
	which impacts the overall compute density of the overall datacenter,
	potentially impacting TCO. \proposed has been carefully designed from the
	ground up to be minimally intrusive to existing server nodes as our
	chiplet-based CPU+FPGA based solution is socket-compatible to current
	systems, easing its adoption.  Third, TensorDIMM is based on a	near-memory
	processing paradigm which requires modifications to the GPU ISA, system
	software, and the runtime system, unlike \proposed which can be implemented
	using existing package-integrated CPU+FPGA technology.  Lastly, TensorDIMM
	relies on rank-level parallelism to increase the effective throughput of
	embedding gather operations, so the benefits of TensorDIMM is limited to
	sufficiently \emph{wide} embedding vectors, constraining the algorithmic
	nature of recommendation models.  Our solution is not tied to a particular
	embedding vector size and is hence much more flexible and applicable for a
	variety of recommendation algorithms.  Overall, the key contribution of our
	work on \proposed is orthogonal to TensorDIMM and stands unique on its own.
	\tab{tab:comparison} is a summary of comparison between \texttt{Centaur} and
	closely related work.

\begin{table*}[t!]
\scriptsize
  \centering
  \caption{Comparison between Centaur and prior work.}
\vspace{-0em}
  \begin{tabular}{|l|c|c|c|c|c|c|c|}
    \hline
     & TABLA~\cite{tabla} &  DNNWEAVER~\cite{dnnweaver} & DNNBuilder~\cite{dnnbuilder} & Cloud-DNN~\cite{cloud_dnn} & Chameleon~\cite{chameleon} & TensorDIMM~\cite{tensordimm} & Ours \\
    \hline
		{\bf Transparent to existing hardware} & \cellcolor{green!25}\Checkmark&\cellcolor{green!25}\Checkmark& \cellcolor{green!25}\Checkmark& \cellcolor{green!25}\Checkmark& & &\cellcolor{green!25}\Checkmark \\
    \hline
		{\bf Transparent to existing software} & \cellcolor{green!25}\Checkmark& \cellcolor{green!25}\Checkmark&\cellcolor{green!25}\Checkmark& \cellcolor{green!25}\Checkmark& & &\cellcolor{green!25}\Checkmark \\

    \hline
		{\bf Work on accelerating dense DNNs}		& \cellcolor{green!25}\Checkmark & \cellcolor{green!25}\Checkmark&\cellcolor{green!25}\Checkmark & \cellcolor{green!25}\Checkmark & & & \cellcolor{green!25}\Checkmark \\
    \hline
		{\bf Applicable for accelerating gathers} & & & & & \cellcolor{green!25}\Checkmark & \cellcolor{green!25}\Checkmark&\cellcolor{green!25}\Checkmark \\
    \hline
		{\bf Applicable for small vector loads} & & & & & \cellcolor{green!25}\Checkmark &  &\cellcolor{green!25}\Checkmark \\
    \hline
		{\bf Study on recommendation models} & & & && & \cellcolor{green!25}\Checkmark&\cellcolor{green!25}\Checkmark \\
    \hline
  \end{tabular}
\vspace{-0.8em}
  \label{tab:comparison}
\end{table*}

\section{Conclusion}
\label{sect:conclusion}

In this paper, we utilize an emerging, package-integrated CPU+FPGA technology
to demonstrate an end-to-end acceleration of personalized recommendation models.
Our hybrid, sparse-dense \proposed architecture synergistically combines a sparse accelerator 
for embedding gathers/reductions and a dense accelerator for GEMM computations,
		holistically addressing the dual challenges of memory bandwidth and compute throughput.
Using a prototype implementation of our proposal on Intel HARPv2 device, \proposed 
achieves $1.7$$-$$17.2\times$ and $1.7$$-$$19.5\times$ performance
and energy-efficiency improvement, respectively, compared to conventional \cpuonly systems.

\section*{Acknowledgment}

This research is supported by Samsung Research Funding Center of Samsung
Electronics (SRFC-TB1703-03). We thank Jaewoong Sim and Intel Labs for giving us
access to the CPU+FPGA system through the Hardware Accelerator Research Program
(HARP). We also thank the anonymous reviewers and Yunjae Lee from our research
group for their constructive feedback that helped improve the final version of
this paper.



\bibliographystyle{IEEEtranS}
\bibliography{ref}

\begin{thebibliography}{10}
\providecommand{\url}[1]{#1}
\csname url@samestyle\endcsname
\providecommand{\newblock}{\relax}
\providecommand{\bibinfo}[2]{#2}
\providecommand{\BIBentrySTDinterwordspacing}{\spaceskip=0pt\relax}
\providecommand{\BIBentryALTinterwordstretchfactor}{4}
\providecommand{\BIBentryALTinterwordspacing}{\spaceskip=\fontdimen2\font plus
\BIBentryALTinterwordstretchfactor\fontdimen3\font minus
  \fontdimen4\font\relax}
\providecommand{\BIBforeignlanguage}[2]{{%
\expandafter\ifx\csname l@#1\endcsname\relax
\typeout{** WARNING: IEEEtranS.bst: No hyphenation pattern has been}%
\typeout{** loaded for the language `#1'. Using the pattern for}%
\typeout{** the default language instead.}%
\else
\language=\csname l@#1\endcsname
\fi
#2}}
\providecommand{\BIBdecl}{\relax}
\BIBdecl

\bibitem{cnvlutin}
J.~Albericio, P.~Judd, T.~Hetherington, T.~Aamodt, N.~E. Jerger, and
  A.~Moshovos, ``{Cnvlutin: Ineffectual-Neuron-Free Deep Convolutional Neural
  Network Computing},'' in \emph{Proceedings of the International Symposium on
  Computer Architecture (ISCA)}, 2016.

\bibitem{mcn}
M.~Alian, S.~W. Min, H.~Asgharimoghaddam, A.~Dhar, D.~K. Wang, T.~Roewer,
  A.~McPadden, O.~O'Halloran, D.~Chen, J.~Xiong, D.~Kim, W.~Hwu, and N.~S. Kim,
  ``{Application-Transparent Near-Memory Processing Architecture with Memory
  Channel Network},'' in \emph{Proceedings of the International Symposium on
  Microarchitecture (MICRO)}, 2018.

\bibitem{altera_fp_ip_core_manual}
Altera, ``{Floating-Point IP Cores User Guide},'' 2016.

\bibitem{alwani:2016:fusedCNN}
M.~Alwani, H.~Chen, M.~Ferdman, and P.~Milder, ``{Fused Layer CNN
  Accelerators},'' in \emph{Proceedings of the International Symposium on
  Microarchitecture (MICRO)}, 2016.

\bibitem{deepspeech_2}
D.~Amodei, R.~Anubhai, E.~Battenberg, C.~Case, J.~Casper, B.~Catanzaro,
  J.~Chen, M.~Chrzanowski, A.~Coates, G.~Diamos, E.~Elsen, J.~Engel, L.~Fan,
  C.~Fougner, T.~Han, A.~Hannun, B.~Jun, P.~LeGresley, L.~Lin, S.~Narang,
  A.~Ng, S.~Ozair, R.~Prenger, J.~Raiman, S.~Satheesh, D.~Seetapun,
  S.~Sengupta, Y.~Wang, Z.~Wang, C.~Wang, B.~Xiao, D.~Yogatama, J.~Zhan, and
  Z.~Zhu, ``{Deep Speech 2: End-To-End Speech Recognition in English and
  Mandarin},'' 2015.

\bibitem{mcm_gpu}
A.~Arunkumar, E.~Bolotin, B.~Cho, U.~Milic, E.~Ebrahimi, O.~Villa, A.~Jaleel,
  C.-J. Wu, and D.~Nellans, ``{MCM-GPU: Multi-Chip-Module GPUs for Continued
  Performance Scalability},'' in \emph{Proceedings of the International
  Symposium on Computer Architecture (ISCA)}, 2017.

\bibitem{chameleon}
H.~Asghari-Moghaddam, Y.~H. Son, J.~H. Ahn, and N.~S. Kim, ``{Chameleon:
  Versatile and Practical Near-DRAM Acceleration Architecture for Large Memory
  Systems},'' in \emph{Proceedings of the International Symposium on
  Microarchitecture (MICRO)}, 2016.

\bibitem{caffe2}
Caffe2, ``{Sparse Operations},'' 2017.

\bibitem{fox}
M.~Campo, C.-K. Hsieh, M.~Nickens, J.~Espinoza, A.~Taliyan, J.~Rieger, J.~Ho,
  and B.~Sherick, ``{Competitive Analysis System for Theatrical Movie Releases
  Based on Movie Trailer Deep Video Representation},'' in \emph{{arxiv.org}},
  2018.

\bibitem{diannao}
T.~Chen, Z.~Du, N.~Sun, J.~Wang, C.~Wu, Y.~Chen, and O.~Temam, ``{DianNao: A
  Small-Footprint High-Throughput Accelerator for Ubiquitous
  Machine-Learning},'' in \emph{Proceedings of the International Conference on
  Architectural Support for Programming Languages and Operation Systems
  (ASPLOS)}, 2014.

\bibitem{eyeriss_isca}
Y.~Chen, J.~Emer, and V.~Sze, ``{Eyeriss: A Spatial Architecture for
  Energy-Efficient Dataflow for Convolutional Neural Networks},'' in
  \emph{Proceedings of the International Symposium on Computer Architecture
  (ISCA)}, 2016.

\bibitem{cloud_dnn}
Y.~Chen, J.~He, X.~Zhang, C.~Hao, and D.~Chen, ``{Cloud-DNN: An Open Framework
  for Mapping DNN Models to Cloud FPGAs},'' in \emph{Proceedings of the
  International Symposium on Field-Programmable Gate Arrays (FPGA)}, 2019.

\bibitem{eyeriss}
Y.~Chen, T.~Krishna, J.~Emer, and V.~Sze, ``{Eyeriss: An Energy-Efficient
  Reconfigurable Accelerator for Deep Convolutional Neural Networks},'' in
  \emph{Proceedings of the International Solid State Circuits Conference
  (ISSCC)}, 2016.

\bibitem{dadiannao}
Y.~Chen, T.~Luo, S.~Liu, S.~Zhang, L.~He, J.~Wang, L.~Li, T.~Chen, Z.~Xu,
  N.~Sun, and O.~Temam, ``{DaDianNao: A Machine-Learning Supercomputer},'' in
  \emph{Proceedings of the International Symposium on Microarchitecture
  (MICRO)}, 2014.

\bibitem{choi:2020:prema}
Y.~Choi and M.~Rhu, ``{PREMA: A Predictive Multi-task Scheduling Algorithm For
  Preemptible Neural Processing Units},'' in \emph{Proceedings of the
  International Symposium on High-Performance Computer Architecture (HPCA)},
  2020.

\bibitem{volta:2017:hotchips}
J.~Choquette, ``{Volta: Programmability and Performance},'' in \emph{Hot Chips:
  A Symposium on High Performance Chips}, 2017.

\bibitem{youtube_recsys}
P.~Covington, J.~Adams, and E.~Sargin, ``{Deep Neural Networks for Youtube
  Recommendations},'' in \emph{Proceedings of the ACM Conference on Recommender
  Systems (RECSYS)}, 2016.

\bibitem{dean:2018:goldenage}
J.~Dean, D.~Patterson, and C.~Young, ``{A New Golden Age in Computer
  Architecture: Empowering the Machine-Learning Revolution},'' in \emph{IEEE
  Micro}, 2018.

\bibitem{bert}
J.~Devlin, M.~Chang, K.~Lee, and K.~Toutanova, ``{BERT: Pre-training of Deep
  Bidirectional Transformers for Language Understanding},'' in
  \emph{{arxiv.org}}, 2018.

\bibitem{fpga:dense5}
C.~Gao, D.~Neil, E.~Ceolini, S.-C. Liu, and T.~Delbruck, ``{DeltaRNN: A
  Power-efficient Recurrent Neural Network Accelerator},'' in \emph{Proceedings
  of the International Symposium on Field-Programmable Gate Arrays (FPGA)},
  2018.

\bibitem{tpu2}
Google, ``{Cloud TPUs: ML Accelerators for TensorFlow},'' 2017.

\bibitem{deeprecsys}
U.~Gupta, S.~Hsia, V.~Saraph, X.~Wang, B.~Reagen, G.-Y. Wei, H.-H.~S. Lee,
  D.~Brooks, and C.-J. Wu, ``{DeepRecSys: A System for Optimizing End-To-End
  At-scale Neural Recommendation Inference},'' in \emph{Proceedings of the
  International Symposium on Computer Architecture (ISCA)}, 2020.

\bibitem{dlrm:arch}
U.~Gupta, C.-J. Wu, X.~Wang, M.~Naumov, B.~Reagen, D.~Brooks, B.~Cottel,
  K.~Hazelwood, M.~Hempstead, B.~Jia, H.-H.~S. Lee, A.~Malevich, D.~Mudigere,
  M.~Smelyanskiy, L.~Xiong, and X.~Zhang, ``{The Architectural Implications of
  Facebook's DNN-based Personalized Recommendation},'' in \emph{Proceedings of
  the International Symposium on High-Performance Computer Architecture
  (HPCA)}, 2020.

\bibitem{song:2015:eie}
S.~Han, X.~Liu, H.~Mao, J.~Pu, A.~Pedram, M.~Horowitz, and W.~J. Dally, ``{EIE:
  Efficient Inference Engine on Compressed Deep Neural Network},'' in
  \emph{Proceedings of the International Symposium on Computer Architecture
  (ISCA)}, 2016.

\bibitem{hazelwood:2018:hpca}
K.~{Hazelwood}, S.~{Bird}, D.~{Brooks}, S.~{Chintala}, U.~{Diril},
  D.~{Dzhulgakov}, M.~{Fawzy}, B.~{Jia}, Y.~{Jia}, A.~{Kalro}, J.~{Law},
  K.~{Lee}, J.~{Lu}, P.~{Noordhuis}, M.~{Smelyanskiy}, L.~{Xiong}, and
  X.~{Wang}, ``{Applied Machine Learning at Facebook: A Datacenter
  Infrastructure Perspective},'' in \emph{Proceedings of the International
  Symposium on High-Performance Computer Architecture (HPCA)}, 2018.

\bibitem{he:www:2017}
X.~He, L.~Liao, H.~Zhang, L.~Nie, X.~Hu, and T.~Chua, ``{Neural Collaborative
  Filtering},'' in \emph{Proceedings of the International Conference on World
  Wide Web (WWW)}, 2017.

\bibitem{hestness:2019:ppopp}
J.~Hestness, N.~Ardalani, and G.~Diamos, ``{Beyond Human-Level Accuracy:
  Computational Challenges in Deep Learning},'' in \emph{Proceedings of the
  Symposium on Principles and Practice of Parallel Programming (PPOPP)}, 2019.

\bibitem{neummu}
B.~Hyun, Y.~Kwon, Y.~Choi, J.~Kim, and M.~Rhu, ``{NeuMMU: Architectural Support
  for Efficient Address Translations in Neural Processing Units},'' in
  \emph{Proceedings of the International Conference on Architectural Support
  for Programming Languages and Operation Systems (ASPLOS)}, 2020.

\bibitem{intel:harpv2}
Intel, ``{Hardware Accelerator Research Program (HARP)},'' 2017.

\bibitem{agilex}
Intel, ``{Intel Agilex FPGAs and SoCs},'' 2019.

\bibitem{intel:foveros}
Intel, ``{Intel Foveros 3D Packaging Technology},'' 2019.

\bibitem{intel:vtune}
Intel, ``{Intel VTune Profiler},'' 2020.

\bibitem{jang:2019:mnnfast}
H.~Jang, J.~Kim, J.-E. Jo, J.~Lee, and J.~Kim, ``{MnnFast: A Fast and Scalable
  System Architecture for Memory-Augmented Neural Networks},'' in
  \emph{Proceedings of the International Symposium on Computer Architecture
  (ISCA)}, 2019.

\bibitem{tpu1}
N.~P. Jouppi, C.~Young, N.~Patil, D.~Patterson, G.~Agrawal, R.~Bajwa, S.~Bates,
  S.~Bhatia, N.~Boden, A.~Borchers, R.~Boyle, P.~luc Cantin, C.~Chao, C.~Clark,
  J.~Coriell, M.~Daley, M.~Dau, J.~Dean, B.~Gelb, T.~V. Ghaemmaghami,
  R.~Gottipati, W.~Gulland, R.~Hagmann, C.~R. Ho, D.~Hogberg, J.~Hu, R.~Hundt,
  D.~Hurt, J.~Ibarz, A.~Jaffey, A.~Jaworski, A.~Kaplan, H.~Khaitan,
  D.~Killebrew, A.~Koch, N.~Kumar, S.~Lacy, J.~Laudon, J.~Law, D.~Le, C.~Leary,
  Z.~Liu, K.~Lucke, A.~Lundin, G.~MacKean, A.~Maggiore, M.~Mahony, K.~Miller,
  R.~Nagarajan, R.~Narayanaswami, R.~Ni, K.~Nix, T.~Norrie, M.~Omernick,
  N.~Penukonda, A.~Phelps, J.~Ross, M.~Ross, A.~Salek, E.~Samadiani, C.~Severn,
  G.~Sizikov, M.~Snelham, J.~Souter, D.~Steinberg, A.~Swing, M.~Tan,
  G.~Thorson, B.~Tian, H.~Toma, E.~Tuttle, V.~Vasudevan, R.~Walter, W.~Wang,
  E.~Wilcox, and D.~H. Yoon, ``{In-Datacenter Performance Analysis of a Tensor
  Processing Unit},'' in \emph{Proceedings of the International Symposium on
  Computer Architecture (ISCA)}, 2017.

\bibitem{snu:ahn:batchnorm}
W.~Jung, D.~Jung, B.~Kim, S.~Lee, W.~Rhee, and J.~Ahn, ``{Restructuring Batch
  Normalization to Accelerate CNN Training},'' in \emph{The Conference on
  Systems and Machine Learning (SysML)}, 2019.

\bibitem{recnmp}
L.~Ke, U.~Gupta, C.-J. Wu, B.~Y. Cho, M.~Hempstead, B.~Reagen, X.~Zhang,
  D.~Brooks, V.~Chandra, U.~Diril, A.~Firoozshahian, K.~Hazelwood, B.~Jia,
  H.-H.~S. Lee, M.~Li, B.~Maher, D.~Mudigere, M.~Naumov, M.~Schatz,
  M.~Smelyanskiy, and X.~Wang, ``{RecNMP: Accelerating Personalized
  Recommendation with Near-Memory Processing},'' in \emph{Proceedings of the
  International Symposium on Computer Architecture (ISCA)}, 2020.

\bibitem{kwon:2019:disagg}
Y.~Kwon and M.~Rhu, ``{A Disaggregated Memory System for Deep Learning},'' in
  \emph{IEEE Micro}, 2019.

\bibitem{tensordimm}
Y.~Kwon, Y.~Lee, and M.~Rhu, ``{TensorDIMM: A Practical Near-Memory Processing
  Architecture for Embeddings and Tensor Operations in Deep Learning},'' in
  \emph{Proceedings of the International Symposium on Microarchitecture
  (MICRO)}, 2019.

\bibitem{mcdla:cal}
Y.~Kwon and M.~Rhu, ``{{A Case for Memory-Centric HPC System Architecture for
  Training Deep Neural Networks}},'' in \emph{IEEE Computer Architecture
  Letters}, 2018.

\bibitem{mcdla}
Y.~Kwon and M.~Rhu, ``{Beyond the Memory Wall: A Case for Memory-Centric HPC
  System for Deep Learning},'' in \emph{Proceedings of the International
  Symposium on Microarchitecture (MICRO)}, 2018.

\bibitem{cambricon}
S.~Liu, Z.~Du, J.~Tao, D.~Han, T.~Luo, Y.~Xie, Y.~Chen, and T.~Chen,
  ``{Cambricon: An Instruction Set Architecture for Neural Networks},'' in
  \emph{Proceedings of the International Symposium on Computer Architecture
  (ISCA)}, 2016.

\bibitem{tabla}
D.~Mahajan, J.~Park, E.~Amaro, H.~Sharma, A.~Yazdanbakhsh, J.~K. Kim, and
  H.~Esmaeilzadeh, ``{TABLA: A Unified Template-based Framework for
  Accelerating Statistical Machine Learning},'' in \emph{Proceedings of the
  International Symposium on High-Performance Computer Architecture (HPCA)},
  2016.

\bibitem{intel:emib}
R.~{Mahajan}, R.~{Sankman}, N.~{Patel}, D.~{Kim}, K.~{Aygun}, Z.~{Qian},
  Y.~{Mekonnen}, I.~{Salama}, S.~{Sharan}, D.~{Iyengar}, and D.~{Mallik},
  ``{Embedded Multi-die Interconnect Bridge (EMIB) -- A High Density, High
  Bandwidth Packaging Interconnect},'' in \emph{IEEE Electronic Components and
  Technology Conference (ECTC)}, 2016.

\bibitem{intel:2018:fpga}
D.~J. Moss, S.~Krishnan, E.~Nurvitadhi, P.~Ratuszniak, C.~Johnson, J.~Sim,
  A.~Mishra, D.~Marr, S.~Subhaschandra, and P.~H. Leong, ``{A Customizable
  Matrix Multiplication Framework for the Intel HARPv2 Xeon+FPGA Platform: A
  Deep Learning Case Study},'' in \emph{Proceedings of the International
  Symposium on Field-Programmable Gate Arrays (FPGA)}, 2018.

\bibitem{intel:2017:fpl}
D.~J. Moss, E.~Nurvitadhi, J.~Sim, A.~Mishra, D.~Marr, S.~Subhaschandra, and
  P.~H. Leong, ``{High Performance Binary Neural Networks on the Xeon+FPGA
  Platform},'' in \emph{Proceedings of the International Conference on Field
  Programmable Logic and Applications (FPL)}, 2017.

\bibitem{facebook_dlrm}
M.~Naumov, D.~Mudigere, H.-J.~M. Shi, J.~Huang, N.~Sundaraman, J.~Park,
  X.~Wang, U.~Gupta, C.-J. Wu, A.~G. Azzolini, D.~Dzhulgakov, A.~Mallevich,
  I.~Cherniavskii, Y.~Lu, R.~Krishnamoorthi, A.~Yu, V.~Kondratenko, S.~Pereira,
  X.~Chen, W.~Chen, V.~Rao, B.~Jia, L.~Xiong, and M.~Smelyanskiy, ``{Deep
  Learning Recommendation Model for Personalization and Recommendation
  Systems},'' in \emph{{arxiv.org}}, 2019.

\bibitem{valgrind}
N.~Nethercote and J.~Seward, ``{Valgrind: A Framework for Heavyweight Dynamic
  Binary Instrumentation},'' in \emph{Proceedings of the ACM SIGPLAN Conference
  on Programming Language Design and Implementation (PLDI)}, 2007.

\bibitem{intel:2017:fpga}
E.~Nurvitadhi, G.~Venkatesh, J.~Sim, D.~Marr, R.~Huang, J.~O.~G. Hock, Y.~T.
  Liew, K.~Srivatsan, D.~Moss, S.~Subhaschandra, and G.~Boudoukh, ``{Can FPGAs
  Beat GPUs in Accelerating Next-Generation Deep Neural Networks?}'' in
  \emph{Proceedings of the International Symposium on Field-Programmable Gate
  Arrays (FPGA)}, 2017.

\bibitem{dgx_1v}
NVIDIA, ``{The NVIDIA DGX-1V Deep Learning System},'' 2017.

\bibitem{volta_v100}
NVIDIA, ``{NVIDIA Tesla V100},'' 2018.

\bibitem{scnn}
A.~Parashar, M.~Rhu, A.~Mukkara, A.~Puglielli, R.~Venkatesan, B.~Khailany,
  J.~Emer, S.~W. Keckler, and W.~J. Dally, ``{SCNN: An Accelerator for
  Compressed-sparse Convolutional Neural Networks},'' in \emph{Proceedings of
  the International Symposium on Computer Architecture (ISCA)}, 2017.

\bibitem{park:2018:olaware}
E.~Park, D.~Kim, and S.~Yoo, ``{Energy-efficient Neural Network Accelerator
  Based on Outlier-aware Low-precision Computation},'' in \emph{Proceedings of
  the International Symposium on Computer Architecture (ISCA)}, 2018.

\bibitem{cosmic}
J.~Park, H.~Sharma, D.~Mahajan, J.~K. Kim, P.~Olds, and H.~Esmaeilzadeh,
  ``{Scale-Out Acceleration for Machine Learning},'' in \emph{Proceedings of
  the International Symposium on Microarchitecture (MICRO)}, 2017.

\bibitem{park:2018:fb}
J.~Park, M.~Naumov, P.~Basu, S.~Deng, A.~Kalaiah, D.~Khudia, J.~Law, P.~Malani,
  A.~Malevich, S.~Nadathur, J.~Pino, M.~Schatz, A.~Sidorov, V.~Sivakumar,
  A.~Tulloch, X.~Wang, Y.~Wu, H.~Yuen, U.~Diril, D.~Dzhulgakov, K.~H.
  an~Bill~Jia, Y.~Jia, L.~Qiao, V.~Rao, N.~Rotem, S.~Yoo, and M.~Smelyanskiy,
  ``{Deep Learning Inference in Facebook Data Centers: Characterization,
  Performance Optimizations and Hardware Implications},'' in
  \emph{{arxiv.org}}, 2018.

\bibitem{rhu:2016:vdnn}
M.~Rhu, N.~Gimelshein, J.~Clemons, A.~Zulfiqar, and S.~W. Keckler, ``{vDNN:
  Virtualized Deep Neural Networks for Scalable, Memory-Efficient Neural
  Network Design},'' in \emph{Proceedings of the International Symposium on
  Microarchitecture (MICRO)}, 2016.

\bibitem{rhu:2018:cdma}
M.~Rhu, M.~O'Connor, N.~Chatterjee, J.~Pool, Y.~Kwon, and S.~W. Keckler,
  ``{Compressing DMA Engine: Leveraging Activation Sparsity for Training Deep
  Neural Networks},'' in \emph{Proceedings of the International Symposium on
  High-Performance Computer Architecture (HPCA)}, 2018.

\bibitem{simba}
Y.~S. Shao, J.~Clemons, R.~Venkatesan, B.~Zimmer, M.~Fojtik, N.~Jiang,
  B.~Keller, A.~Klinefelter, N.~Pinckney, P.~Raina, S.~G. Tell, Y.~Zhang, W.~J.
  Dally, J.~Emer, C.~T. Gray, B.~Khailany, and S.~W. Keckler, ``{Simba: Scaling
  Deep-Learning Inference with Multi-Chip-Module-Based Architecture},'' in
  \emph{Proceedings of the International Symposium on Microarchitecture
  (MICRO)}, 2019.

\bibitem{dnnweaver}
H.~Sharma, J.~Park, D.~Mahajan, E.~Amaro, J.~K. Kim, C.~Shao, A.~Misra, and
  H.~Esmaeilzadeh, ``{From High-Level Deep Neural Models to FPGAs},'' in
  \emph{Proceedings of the International Symposium on Microarchitecture
  (MICRO)}, 2016.

\bibitem{fpga:dense1}
Y.~Shen, M.~Ferdman, and P.~Milder, ``{Maximizing CNN Accelerator Efficiency
  Through Resource Partitioning},'' in \emph{Proceedings of the International
  Symposium on Computer Architecture (ISCA)}, 2017.

\bibitem{kdd:alibaba}
J.~Wang, P.~Huang, H.~Zhao, Z.~Zhang, B.~Zhao, and D.~L. Lee, ``{Billion-scale
  Commodity Embedding for E-commerce Recommendation in Alibaba},'' in
  \emph{Proceedings of the International Conference on Knowledge Discovery and
  Data Mining (KDD)}, 2018.

\bibitem{whatmough:2017:hotchips}
P.~N. Whatmough, S.~K. Lee, N.~Mulholland, P.~Hansen, S.~Kodali, D.~C. Brooks,
  and G.-Y. Wei, ``{DNN ENGINE: A 16nm Sub-uJ Deep Neural Network Inference
  Accelerator for the Embedded Masses},'' in \emph{Hot Chips: A Symposium on
  High Performance Chips}, 2017.

\bibitem{wu:2019:fb_edge}
C.-J. Wu, D.~Brooks, K.~Chen, D.~Chen, S.~Choudhury, M.~Dukhan, K.~Hazelwood,
  E.~Isaac, Y.~Jia, B.~Jia, T.~Leyvand, H.~Lu, Y.~Lu, L.~Qiao, B.~Reagen,
  J.~Spisak, F.~Sun, A.~Tulloch, P.~Vajda, X.~Wang, Y.~Wang, B.~Wasti, Y.~Wu,
  R.~Xian, S.~Yoo, and P.~Zhang, ``{Machine Learning at Facebook: Understanding
  Inference at the Edge},'' in \emph{Proceedings of the International Symposium
  on High-Performance Computer Architecture (HPCA)}, 2019.

\bibitem{sigarch:blog:recsys}
C.-J. Wu, D.~Brooks, U.~Gupta, H.-H. Lee, and K.~Hazelwood, ``{Deep Learning:
  It’s Not All About Recognizing Cats and Dogs},'' 2019.

\bibitem{fpga:dense3}
Q.~{Xiao}, Y.~{Liang}, L.~{Lu}, S.~{Yan}, and Y.~{Tai}, ``{Exploring
  Heterogeneous Algorithms for Accelerating Deep Convolutional Neural Networks
  on FPGAs},'' in \emph{Design Automation Conference (DAC)}, 2017.

\bibitem{fpga:dense4}
C.~Zhang, P.~Li, G.~Sun, Y.~Guan, B.~Xiao, and J.~Cong, ``{Optimizing
  FPGA-based Accelerator Design for Deep Convolutional Neural Networks},'' in
  \emph{Proceedings of the International Symposium on Field-Programmable Gate
  Arrays (FPGA)}, 2015.

\bibitem{fpga:dense2}
J.~Zhang and J.~Li, ``{Improving the Performance of OpenCL-based FPGA
  Accelerator for Convolutional Neural Network},'' in \emph{Proceedings of the
  International Symposium on Field-Programmable Gate Arrays (FPGA)}, 2017.

\bibitem{dnnbuilder}
X.~Zhang, J.~Wang, C.~Zhu, Y.~Lin, J.~Xiong, W.~Hwu, and D.~Chen,
  ``{DNNBuilder: An Automated Tool for Building High-Performance DNN Hardware
  Accelerators for FPGAs},'' in \emph{Proceedings of the International
  Conference on Computer-Aided Design (ICCAD)}, 2018.

\end{thebibliography}

\end{document}